\documentclass[longauth]{aa} 
\usepackage{graphicx}
\usepackage{txfonts}
\usepackage{rotating}
\usepackage{xcolor}
\usepackage[hidelinks]{hyperref}
\usepackage{mathtools}

\pdfoutput=1

\begin{document} 

\title{The miniJPAS survey:}
\subtitle{Optical detection of galaxy clusters with PZWav}
\authorrunning{L. Doubrawa and J-PAS collaboration}
\titlerunning{PZWav-miniJPAS detection of galaxy clusters}


\author{
    L. Doubrawa\inst{\ref{astrosaopaulo},\ref{helsinki}}
    \and
    E. S. Cypriano\inst{\ref{astrosaopaulo}}
    \and
    A. Finoguenov\inst{\ref{helsinki}}
    \and
    P. A. A. Lopes\inst{\ref{OV}}
    \and
    A. H. Gonzalez\inst{\ref{florida}}
    \and
    M. Maturi\inst{\ref{germany1},\ref{germany2}}
    \and    
    R. A. Dupke\inst{\ref{ON},\ref{DAUM},\ref{DPA}}
    \and
    R.~M.~Gonz\'alez Delgado\inst{\ref{iaa}}  
    \and
    R.~Abramo\inst{\ref{fisicasaopaulo}} 
    \and
    N.~Benitez\inst{\ref{iaa}} 
    \and    
    S. Bonoli\inst{\ref{teruel},\ref{DIPC},\ref{Ikerb}}  
    \and
    S.~Carneiro\inst{\ref{fisicabahia}}  
    \and
    J.~Cenarro\inst{\ref{teruel}}  
    \and
    D.~Cristóbal-Hornillos\inst{\ref{teruel}}  
    \and
    A.~Ederoclite\inst{\ref{teruel}}  
    \and
    A.~Hern\'an-Caballero\inst{\ref{teruel}} 
    \and
    C. López-Sanjuan\inst{\ref{teruel}} 
    \and
    A. Marín-Franch\inst{\ref{teruel}}  
    \and
    C. Mendes~de~Oliveira\inst{\ref{astrosaopaulo}}  
    \and
    M. Moles\inst{\ref{teruel}}  
    \and
    L. Sodré~Jr.\inst{\ref{astrosaopaulo}}  
    \and
    K. Taylor\inst{\ref{instruments4}}  
    \and
    J. Varela\inst{\ref{teruel}}  
    \and
    H. V\'azquez Rami\'o\inst{\ref{teruel}} 
}

\institute{
    Departamento de Astronomia, Instituto de Astronomia, Geof\'isica e Ci\^encias Atmosf\'ericas da USP, Cidade Universit\'aria, \\ 05508-900, S\~ao Paulo, SP, Brazil\label{astrosaopaulo}
    \email{lia.doubrawa@usp.br}
\and
    Department of Physics, University of Helsinki, P.O. Box 64, FI-00014 Helsinki, Finland \label{helsinki}
\and
    Observat\'orio do Valongo, Universidade Federal do Rio de Janeiro, Ladeira do Pedro Ant\^onio 43, Rio de Janeiro, RJ, 20080-090, Brazil\label{OV}
\and
    Department of Astronomy, University of Florida, Gainesville, FL 32611-2055, USA  \label{florida}
\and
    Zentrum f\"ur Astronomie, Universitat\"at Heidelberg, Philosophenweg 12, D-69120 Heidelberg, Germany \label{germany1}
\and
    Institute for Theoretical Physics, Philosophenweg 16, D-69120 Heidelberg, Germany  \label{germany2}
\and
    Observatório Nacional, Rua General José Cristino, 77, São Cristóvão, 20921-400, Rio de Janeiro, RJ, Brazil\label{ON}
\and
    Department of Astronomy, University of Michigan, 311 West Hall, 1085 South University Ave., Ann Arbor, USA\label{DAUM}
\and
    Department of Physics and Astronomy, University of Alabama, Box 870324, Tuscaloosa, AL, USA\label{DPA}
\and
    Instituto de Astrof\'isica de Andaluc\'ia (IAA-CSIC), Glorieta de la Astronomía s/n, E-18008 Granada, Spain\label{iaa}
\and
    Departamento de F\'isica Matem\'atica, Instituto de F\'{\i}sica, Universidade de S\~ao Paulo, Rua do Mat\~ao, 1371, CEP 05508-090, S\~ao Paulo, Brazil \label{fisicasaopaulo}
\and
    Centro de Estudios de F\'isica del Cosmos de Arag\'on (CEFCA), Unidad Asociada al CSIC, Plaza San Juan, 1, E-44001 Teruel, Spain\label{teruel}
\and
    Donostia International Physics Center (DIPC), Manuel Lardizabal Ibilbidea, 4, San Sebasti\'an, Spain\label{DIPC}
\and
    Ikerbasque, Basque Foundation for Science, 48013 Bilbao, Spain\label{Ikerb}
\and
    Instituto de F\'isica, Universidade Federal da Bahia, 40210-340, Salvador, BA, Brazil\label{fisicabahia}
\and
    Instruments4, 4121 Pembury Place, La Canada Flintridge, CA 91011, U.S.A\label{instruments4}
}

\date{\today}

\abstract
   {Galaxy clusters are an essential tool to understand and constrain the cosmological parameters of our Universe. Thanks to its multi-band design, J-PAS offers a unique group and cluster detection window using precise photometric redshifts and sufficient depths.   }
   {We produce galaxy cluster catalogues from the miniJPAS, which is a pathfinder survey for the wider J-PAS survey, using the PZWav algorithm.}
   {Relying only on photometric information, we provide optical mass tracers for the identified clusters, including richness, optical luminosity, and stellar mass. By reanalysing the Chandra mosaic of the AEGIS field, alongside the overlapping XMM-Newton observations, we produce an X-ray catalogue.}
   {The analysis reveals the possible presence of structures with masses of $4\times 10^{13}$ M$_\odot$ at redshift $0.75$, highlighting the depth of the survey. Comparing results with those from two other cluster catalogues, provided by AMICO and VT, we find $43$ common clusters with cluster centre offsets of $100\pm60$ kpc and redshift differences below $0.001$. We provide a comparison of the cluster catalogues with a catalogue of massive galaxies and report on the significance of cluster selection. In general, we are able to recover approximately $75\%$ of the galaxies with $M^{\star} > 2 \times 10^{11}$ M$_\odot$.}
   {This study emphasises the potential of the J-PAS survey and the employed techniques down to the group scales.}


\maketitle

\section{Introduction}

Galaxy clusters are powerful tracers of cosmic structure formation.  Linked to massive overdensity peaks in the Universe, cluster abundance as a function of mass and redshift provides a compelling tool to constrain the cosmological parameters \citep[e.g.][]{Carlberg1996, Reiprich2002, Voit2005, Allen2011, Weinberg2013, Pacaud2016, Costanzi2019, Chitham2020, Giocoli2021, Lesci2022, Lesci2022A}. Thus, the identification and characterisation of galaxy clusters in different wavelengths and numerical simulations allow us to refine our understanding of the Universe's structure, evolution and fundamental physics.

Wide-field cosmological surveys, such as DES \citep{des2005}, KiDS \citep{deJong2013}, J-PAS \citep{Benitez2014} and in the near future Euclid \citep{Euclid2022} provide information about billions of galaxies. Of particular interest to group and galaxy cluster detection, sophisticated algorithms and data analysis techniques have been developed to identify and catalogue large-scale structures and characterise physical properties such as mass, morphology, stellar mass content, and gas properties. These algorithms employ diverse methodologies, including detecting overdensities in galaxy distributions, analysing multi-wavelength data, and machine learning techniques. In this work, we focus on a specific cluster finder, the adaptive wavelet filtering technique PZWav \citep{Gonzalez14, Werner2022}, and compare results with those from the Voronoi tessellation method, VT \citep{Ramella2001, Lopes2004} and the matched filtering algorithm, AMICO \citep[][]{Bellagamba2019, Maturi2019}. 
Cluster finders such as AMICO and PZWav have been developed with the ability to exploit the photometric information in the form of a redshift probability density function (PDF). This feature proves particularly valuable in the recent imaging surveys that utilise narrow-band filters \citep[e.g.][]{Benitez2014, Bonoli2021, Marti2014, Mendes2019, Cenarro2019} as it enables harnessing the precise photometric redshift estimates of galaxies.

In this study, we utilise the accurate photometric redshift estimates obtained from the miniJPAS survey (see details in \S 2) to assess the strengths of the recently started wide J-PAS survey. Our analysis focuses on testing the performance of the PZWav cluster finder and its comparison to the other two methods (VT and AMICO), extending the limits to the regime of low-mass groups. 
As the miniJPAS data set also provides reliable spectral energy distribution (SED) estimates \citep{Delgado2021, Delgado2022}, we characterise the identified structures by evaluating richness, optical luminosity, and total stellar mass.
The paper is organised as follows. In \S\,\ref{Data} we describe the photometric information obtained from the miniJPAS survey. \S\,\ref{Methods} presents details of the chosen algorithm to detect the galaxy cluster candidates and introduces the adaptive membership estimator utilised to characterise the cluster sample. In \S\,\ref{Catalogue}, we outline the galaxy cluster catalogue, propose two absolute magnitude regimes, and discuss a richness selection. In \S\,\ref{Results}, we show the matching cluster candidates within the overlapping AEGIS\footnote{All-Wavelength Extended Groth Strip International Survey} X-ray field, we provide scaling relations given the mass tracers, and we discuss the properties of catalogues obtained from different algorithms. We also present the significance of our results based on a catalogue of massive galaxies. We summarise and conclude in \S\,\ref{Conclusions}.
Through this work, we adopt a flat $\Lambda$CDM cosmology, with $H_{0}=68, \Omega_{m}=0.31$, in agreement to \citet{Planck2016} parameters. Magnitudes are given in the AB system.

\section{Data} \label{Data}

We utilise the photometric data obtained by \cite{Hernan2021} from the miniJPAS survey \citep{Bonoli2021}, which comprises a set of $54$ narrow-band (FWHM$\sim145$\AA, following the SDSS broad-band filters $u,g,r,i$), along with two broad-band filters, extending to the near-infrared and the near-UV. This survey covers an approximate area of $1$ squared degree on the Extended Groth Strip field. 
The miniJPAS data set served as a pathfinder for the recently started wide-field Cosmological Survey, the Javalambre-Physics of the Accelerated Universe Astrophysical Survey (J-PAS). Conducted from the Javalambre Observatory in Spain, the survey employs a dedicated $2.5$m telescope and a $1.2$ Gpix camera capable of capturing in a single-shot an area of $4.7$ square degrees of the sky \citep{Benitez2014}. The complete survey aims to observe $8500$ squared degrees using the filter system described above.     

The $z_{phot}$ were estimated using a customised version of the LePhare code \citep{Arnouts2011}, which has been specifically optimised to replicate the J-PAS filter system. 
This code computes the $z_{phot}$ probability density function (PDF) by weighting the log-likelihood distribution with a prior that accounts for the redshift distribution, considering the galaxies' magnitude and colour. The prior function is derived from the galaxy spectroscopic redshift distribution obtained by the VIMOS VLT Deep Survey \citep{LeFevre2005}, which also provides the redshift and spectral type probability density functions in the $i$-band magnitude.

To understand and characterise the $z_{phot}$ performance, \cite{Hernan2021} and \cite{Laur2022L} analysed sub-samples of galaxies with available spectroscopic redshifts in the miniJPAS catalogue. The study performed by \cite{Hernan2021} revealed a typical $z_{phot}$ uncertainty of $\sigma_{MAD,z}=0.013$ for galaxies with magnitude in $r$-band $r<23$. 
As a subsequent work, \cite{Laur2022L} present a new $z_{phot}$  workflow to provide redshift estimates for the J-PAS survey.

This first data release from the miniJPAS provides an ideal data set for refining various strategies related to galaxy cluster detection and analysis, offering consistent insights for the J-PAS survey.

\section{Methods} \label{Methods}

This section provides details regarding the PZWav detection algorithm and the run parameters utilised in this analysis. Additionally, we introduce a probabilistic approach to characterise the cluster sample, focusing on identifying potential galaxy members within a characteristic radius.

\subsection{PZWav algorithm} \label{pzwav_description}

There are a myriad of algorithms for the detection of galaxy clusters using optical data \citep[for a review][]{Roy2006, Euclid2019III}. Among them, PZWav \citep{Gonzalez14, Werner2022} and AMICO \citep{Bellagamba2019, Maturi2019} are particularly suitable for the J-PAS due to their capacity to handle not only the nominal values of the photometric redshifts, but also their PDFs. PZWav is a density-based algorithm that requires minimal assumptions about cluster properties, making it complementary to matched-filter approaches such as the one implemented in AMICO, for instance.

PZWav performs the detection of structures by identifying overdensities on fixed physical scales associated with clusters. It requires information such as sky coordinates of galaxies, photometric redshifts, and magnitudes. The algorithm creates a series of redshift slices, with each galaxy assigned a weight based on the probability of its redshift lying at the respective bin. By integrating the probabilities over the redshift limits of each bin, the code generates 2D galaxy density maps. These maps are convoluted with a difference-of-Gaussians kernel, which enhances cluster-sized structures while diminishing the impact of smaller and larger structures. Additionally, to estimate the related noise, a separate set of density maps is created by randomising the projected positions of galaxies within the redshift slice.\footnote{This is slightly different than the approach used for MaDCoWS2 (Thongkham et al, in prep), where the positions are preserved but the PDFs are shuffled.}

In each redshift slice, a galaxy cluster candidate is identified as a peak in the density map that exceeds the noise threshold. The cluster centre is defined as the position of the local highest peak, and the final redshift of the cluster is determined by computing the median $z_{phot}$ of the galaxies within a redshift range $\Delta z = 0.02$ (equivalent to twice the bin width) and a physical radius of $R=500$ kpc. To identify the structures, we define redshift slices with a width of $dz_w=0.01$ and constrain the cluster smoothing kernel scales to $400$ and $1400$ kpc. 

A signal-to-noise ratio (SNR) is defined as the amplitude of the highest peak in the density maps in relation to the noise level. The noise level is calculated as the standard deviation of a Gaussian approximation obtained from density maps produced by a randomised set of coordinates for the galaxy positions. 
For this analysis, the default detection threshold is set to SNR$_{\rm thr}=4$. Tests with simulations, as performed by \cite{Werner2022}, indicates that our catalogue is expected to be $85\%$ pure and $80\%$ complete\footnote{In \cite{Werner2022}, the authors use a prior version of PZWav, with slightly different key parameters used to optimise the performance for the S-PLUS survey. Despite the slightly different configuration, we find similar values for purity and completeness. The noise calculation is similar to the method used in our work.}

To prevent double counting of clusters passing through a collision event, we set as merging parameters $dr_{lim}=1500$ kpc and $dz_m=0.03$ as the minimum distance between two structures in both plane-of-the-sky and redshift spaces. 

In our study, we consider all galaxies brighter than $m = 22.7$ in the $r$-band. No cuts or selection criteria are applied based on $z_{phot}$ quality, colour, or other factors. This approach allows us to analyse the full galaxy sample, providing a complete view of the cluster population.

\subsection{An adaptive membership estimator (AME)} \label{Richness_code}

For further characterisation of the cluster detection catalogue, we account with a probabilistic approach to estimate galaxy memberships. While spectroscopic surveys ideally provide accurate membership information, their demanding observational requirements make achieving high coverage fractions of cluster galaxies challenging.
Therefore, the availability of photometric surveys encourages the development of strategies that rely on photometric information to estimate galaxy memberships.

The detection algorithm PZWav does not provide a galaxy member catalogue alongside the cluster candidate detection. Therefore, we have developed an adaptive membership estimator (AME) that is density-based, and aims to be as much data-based and assumptions-free as possible. AME utilises the cluster information in the (2+1)D space, specifically the cluster redshift (z$_{cl}$), sky positions (R.A.$_{cl}$ Dec$_{cl}$), and the galaxies $z_{phot}$ and PDFs.
For a detailed review of the method and its application on a simulated data set, see \cite{Doubrawa2023}.

Below, we summarise the relevant steps of the algorithm.

First, we exclude obvious non-members by selecting the galaxies within a radius of $2.5$ Mpc in the plane of the sky from the cluster centre, and with $|z_{phot} - z_{cl}| > 3\,\sigma_{MAD,z}(1+z_{cl})$. Where $\sigma_{MAD,z}$ is the photo-$z$ error, as presented in \S\,\ref{Data} \citep[$\sigma_{MAD,z}=0.013$, ][]{Hernan2021}.

Then, we define an aperture radius ($R_c$) characterised by a change of a factor of $2$ in the radial density profile of galaxies. This parameter highlights the change from cluster to field-dominated galaxies.

We then shuffle the redshift of each galaxy within $R_c$ by randomly assigning a value based on the galaxy photo-$z$ PDFs ($z_{PDF}$). We apply a $3\sigma$ clipping on the velocity dispersion estimated with $z_{PDF}$ to avoid contamination by field galaxies.

With the remaining galaxies, we apply HDBSCAN\footnote{Hierarchical Density-Based Spatial Clustering of Applications with Noise.} \citep{Campello2014}, a density-based clustering algorithm that connects galaxies that are spatially close, considering the isolated ones as interlopers. This step identifies the largest group of galaxies linked by HDBSCAN as cluster members.

To use the full photo-$z$’s PDF information, we repeat the last two steps $100$ times. The final membership probability (P$_{\text{mem}}$) for a given galaxy is then defined as the number of times it is selected as a member galaxy over the total number of iterations.

\subsection{Cluster characterisation}

One of J-PAS's, and also miniJPAS's, main attributes is the high precision achieved through the filter system. The characterisation and template fittings, incorporating colour and galaxy extinction information, result in accurate measurements of absolute magnitudes and stellar mass \citep{Delgado2021, Delgado2022}.
Exploring these attributes, we characterise the cluster sample by estimating richness, total optical luminosity, and total stellar mass within $R_c$. We present the results in two different absolute magnitude limits ($-19.5$ and $-21.25$), as presented in detail in \S\,\ref{sec:groups_cls}.  

We define richness as the sum of the probabilities of all galaxies associated with a given cluster, calculated as $\lambda = \sum_i\, P_{\text{mem},i}$. Similarly, the total optical luminosity is determined by summing the luminosity of each galaxy, $L_i = 10^{0.4[4.65 - R_i]}$, weighted by the galaxy pertinence, $L_\lambda=\sum L_i\, P_{\text{mem},i}$. Here, $4.65$ represents the solar absolute magnitude in $r$-band \citep{Willmer2018}, and $R_i$ is the galaxy absolute magnitude in the same band. Additionally, we can derive the total stellar mass of the cluster sample by summing the stellar masses of individual galaxies ($M_i^{\star}$) weighted by their membership probabilities, resulting in $M^{\star} = \sum_i M_i^{\star} P_{\text{mem},i}$.

\section{PZWav's Galaxy cluster catalogue} \label{Catalogue}

In this section, we present the properties of the resulting galaxy cluster catalogue. We examine the impact of applying different absolute magnitude limits focusing on the completeness of the catalogue at high redshifts and a better characterisation of clusters at low redshifts. Furthermore, we implement a richness selection to improve the catalogue purity by removing possible detections by chance.

\subsection{Catalogue} \label{sec:catalogue}

\begin{table}
    \caption{Galaxy cluster detection catalogue columns and description.}
    \centering
    \begin{tabular}{l l}
        \hline
         ID  & PZWav cluster identification number\\
         RA  & PZWav position in the sky in R.A.\\
         DEC & PZWav position in the sky in Dec.\\
         z   & PZWav redshift \\
         SNR & PZWav signal-to-noise \\
         
         $\lambda$-19.5 & Richness estimates (M$_r$<-19.5)\\
         $\lambda$\_err-19.5 & Uncertainty in $\lambda$ (M$_r$<-19.5)\\         
         Lum-195 & Optical luminosity (M$_r$<-19.5) [L$_\odot$]\\
         Lum\_err-19.5 & Uncertainty in $L_\lambda$  (M$_r$<-19.5) [L$_\odot$]\\
         Mstar-19.5 & Stellar mass (M$_r$<-19.5) [M$_\odot$]\\
         Mstar\_err-19.5 & Uncertainty in $M_\lambda^\star$  (M$_r$<-19.5) [M$_\odot$]\\
         
         $\lambda$-21.25 & Richness estimates (M$_r$<-21.25) \\
         $\lambda$\_err-21.25 & Uncertainty in $\lambda$  (M$_r$<-21.25) \\
         Lum-21.25 & Optical luminosity (M$_r$<-21.25) [L$_\odot$]\\
         Lum\_err-21.25 & Uncertainty in $L_\lambda$ (M$_r$<-21.25) [L$_\odot$]\\
         Mstar-21.25 & Stellar mass (M$_r$<-21.25) [M$_\odot$]\\
         Mstar\_err-21.25 & Uncertainty in $M_\lambda^\star$ (M$_r$<-21.25) [M$_\odot$]\\
         
         \hline
    \end{tabular}

    \label{tab:pzwav_columns}
\end{table}

Running the PZWav algorithm on the miniJPAS area, we obtain a catalogue with $574$ galaxy groups and cluster candidates in the redshift range of $0.05 < z < 0.8$, above an SNR of $4$. The catalogue provides essential information, including the identification number, sky coordinates, photometric redshift, and detected SNR of each cluster. Additional properties such as richness, total stellar mass and optical luminosity are given for two different absolute magnitude ranges (as discussed in \S\,\ref{sec:groups_cls}). 
Table\,\ref{tab:pzwav_columns} provides a summary of the available columns in the catalogue along with their corresponding descriptions.

\begin{figure*}
    \centering
    \includegraphics[width=0.97\columnwidth]{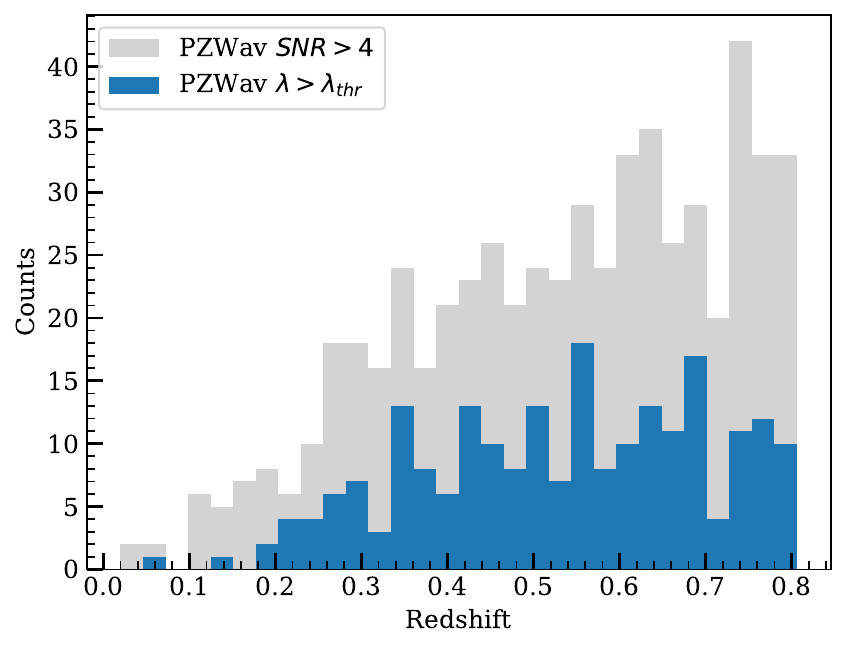}
    \includegraphics[width=\columnwidth]{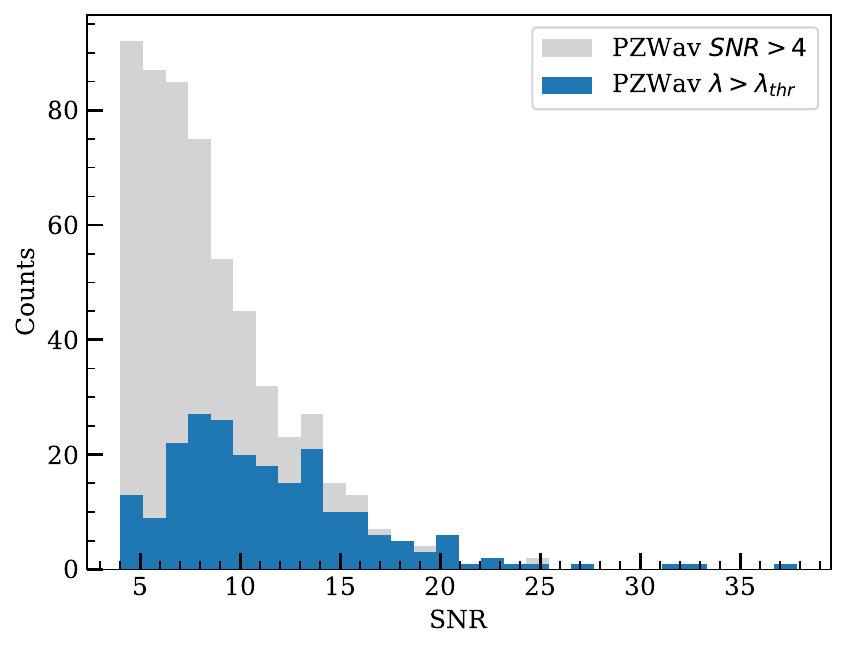}

    \caption{Left panel: Redshift distribution of the cluster candidates detected by PZWav. Right panel: PZWav SNR distribution. Both histograms show the sample before (grey) and after (blue bars) the richness selection. We apply $\lambda_{thr}$ of $1.9$ and $3.0$ for cluster and group regimes (as discussed in \S\ref{sec:groups_cls}).}
    \label{fig:pzwav_detections}
\end{figure*}

In Figure\,\ref{fig:pzwav_detections}, we present the catalogue's redshift (top panel) and SNR distributions in histograms (bottom panel). We highlight two distinct samples: the complete catalogue, represented by the grey bars, which includes all clusters detected above the SNR threshold of 4, and the clean sample (discussed in \S\,\ref{sec:decontamination}), indicated by the blue bars.

\subsection{Cluster member catalogue} \label{sec:membership}

Understanding the membership composition of galaxy clusters is crucial for a comprehensive analysis of their properties and dynamics. In this section, we present the membership catalogue produced by AME (\S\ref{Richness_code}), a fundamental component of our study, which allows us to identify and characterise the galaxies that belong to the detected galaxy clusters. 

The catalogue comprises $9070$ galaxy members, which accounts for $82\%$ of the available galaxy catalogue.
Table \ref{tab:ame_output} provides an overview of the key information included in the membership catalogue. This data set contains essential data, such as sky positions, photometric redshift, photo-z PDF, and membership probability. $P_{\rm mem}$ is listed for two absolute magnitude limits, which will be further discussed in detail in the following section (\S\ref{sec:groups_cls}). 

\begin{table}
    \caption{Galaxy membership catalogue columns and description. }  
    \centering
    \begin{tabular}{l l}
        \hline
         ID  & Identification number\\
         RA  & Position in the sky in R.A.\\
         DEC & Position in the sky in Dec.\\
         photo\_z & Photometric redshift \\ 
         MAG\_AUTO & Apparent magnitude in $r$-band\\ 
         ClusterID & Corresponding Cluster ID\\
         Pmem-19.5 & Membership probability for M$_r$<-19.5\\
         Pmem-21.25 & Membership probability for M$_r$<-21.25\\
         \hline
    \end{tabular}
  \label{tab:ame_output}
\end{table}

A detailed analysis of the membership significance and statistics can be found in \cite{Doubrawa2023}. In addition to the galaxy catalogue obtained by the PZWav algorithm, we have extended our analysis to include membership catalogues for the diverse cluster catalogues investigated throughout this study. The catalogues can be found at the J-PAS website.

\subsection{Group and cluster regimes} \label{sec:groups_cls}

To avoid systematic errors in richness, total luminosity, and stellar mass estimations, such as the incompleteness of the galaxy catalogue due to the redshift range, we chose to work in two different regimes:
\begin{itemize}
    \item  The {\bf cluster regime}, on which we consider galaxies with an absolute magnitude below M$_r< -21.25$, ensuring a volume-limited sample up to a redshift of $z < 0.8$. 
    \item The {\bf group regime}, with an absolute magnitude limit of M$_r< -19.5$, valid for $z < 0.3$.
\end{itemize}

Considering their different luminosity properties, these regimes allow us to study structures with a different focus, with a brighter cut benefiting galaxy group-like structures. 

\begin{figure*}
    \centering
    \includegraphics[width=\columnwidth]{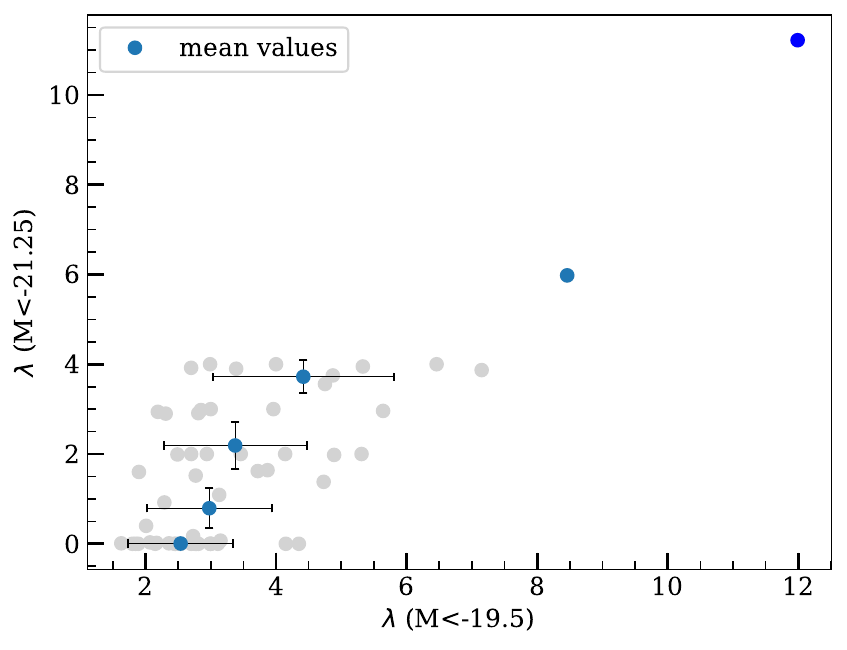}  
    \includegraphics[width=\columnwidth]{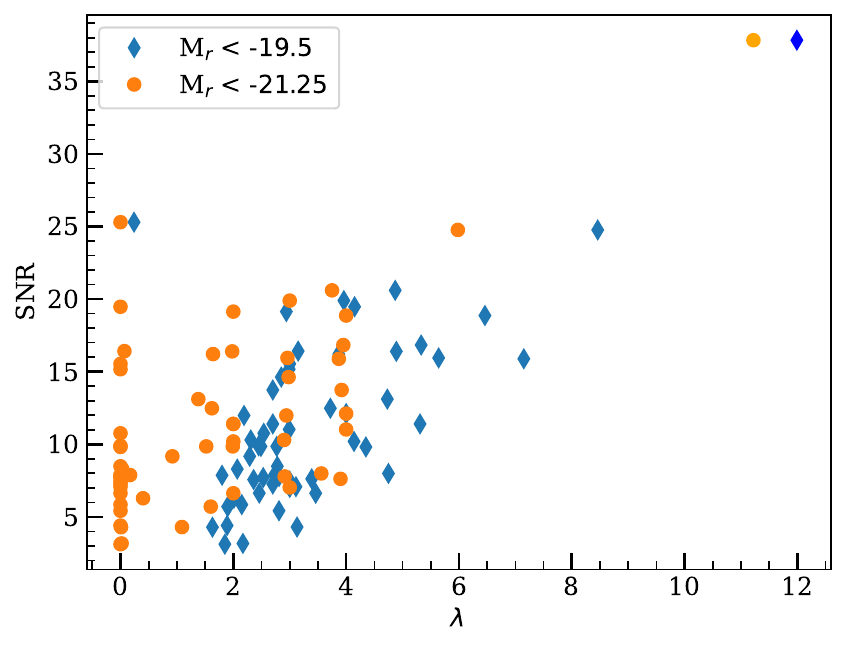}
    \caption{Left panel: Comparison within richness calculated in two different absolute magnitude cuts: group regime (M$_r<-19.5$) for structures with $z<0.3$, and cluster regime (M$_r<-21.25$) within $0.3<z<0.8$. Right panel: PZWav $SNR$ in the function of the derived richness. Some groups overlooked in the cluster regime can have a significant richness in shallower magnitude cuts.}
    \label{fig:comp_rich}
\end{figure*}

In the left panel of Figure\,\ref{fig:comp_rich}, we present the richness estimates for the galaxy cluster candidates concentrated within the redshift range of $0.05 < z < 0.3$. We compare the richness estimates obtained with the two absolute magnitude cuts of M$_r < -21.25$ (cluster regime) and M$_r < -19.5$ (group regime). The blue points in the left panel show the median richness values, with error bars representing the dispersion of the sample. The figure highlights the larger values in richness estimation when considering fainter magnitudes. For instance, systems that are overlooked with a richness of $\lambda = 0$ using the bright magnitude cut increase to $\lambda = 4.2$ when fainter galaxies are included. 

The right panel of the figure focuses on SNR provided by PZWav for the selected candidates. It is important to note that we do not apply magnitude cuts while running the detection code. However, while analysing the detected structures in the cluster regime, some of them exhibit a richness value of $\lambda = 0$, and still show a high SNR. This indicates that these candidates have a high-density amplitude compared to the background noise and were not adequately described by the selected bright absolute magnitude cut. However, when the group regime is applied, the richness values are correctly estimated, allowing a more accurate characterisation.

This test emphasises the importance of considering different magnitude ranges depending on the desired significance level.

\subsection{Generating a refined catalogue using richness}\label{sec:decontamination} 

One of the challenges in galaxy cluster detections is assessing the contamination and completeness of the resulting catalogue. The presence of false structures can introduce biases in the measurements of cluster abundances, thus affecting our understanding of the underlying cosmological parameters based, for example, on counts and clustering. Studies with simulated sky areas presented by \cite{Werner2022} for the S-PLUS survey \citep{Mendes2019} have shown that PZWav SNR$ > 3.3$ achieves the best values of completeness and purity. In contrast, other approaches, such as proposed by \cite{Rykoff2014}, adopt a distinct selection method, using a richness threshold of $20$ galaxies to limit their sample for higher mass clusters. 

Here we propose an alternative approach to address contamination in the catalogue. Following a similar methodology described in \cite{Klein_2017}, we develop a selection criteria based on the richness that allows us to remove cluster candidates below a certain threshold.

The method involves the comparison of the PZWav catalogue with randomly distributed points in both sky positions and redshifts. We expect to find small richness values, as these sky positions are not privileged, but sometimes, this distribution may reach a galaxy overdensity and provide a meaningful measurement.
For each PZWav cluster candidate, we evaluate the number of random points within a specific redshift bin defined by $|z_{opt, i}-z_{rand}|<0.05$. We then calculate the fraction of these points with a richness value lower than the richness of the $i$-th PZWav candidate. This process provides a quantitative measurement of the significance of the calculated richness compared to the randomly distributed points.
By setting a significance threshold of $90\%$, we can identify and remove cluster candidates with potential detection by chance given a richness value.

To minimise the possibility of misidentifying a galaxy group as contamination due to the magnitude cut (as presented in \S\,\ref{sec:groups_cls}), we repeat the above analysis in the two magnitude regimes and obtain two different richness values. Therefore, to ensure the $90\%$ level of significance we establish richness limits of $1.9$ and $3.0$ for cluster and group regimes. As a result, the final content of the catalogue is reduced to only $38\%$ of the initial cluster candidate sample. The impact of this cutoff is evident in the altered distributions of redshift and SNR, as illustrated by the blue bars in Figure\,\ref{fig:pzwav_detections}. 

\begin{figure}
    \centering
    \includegraphics[width=\columnwidth]{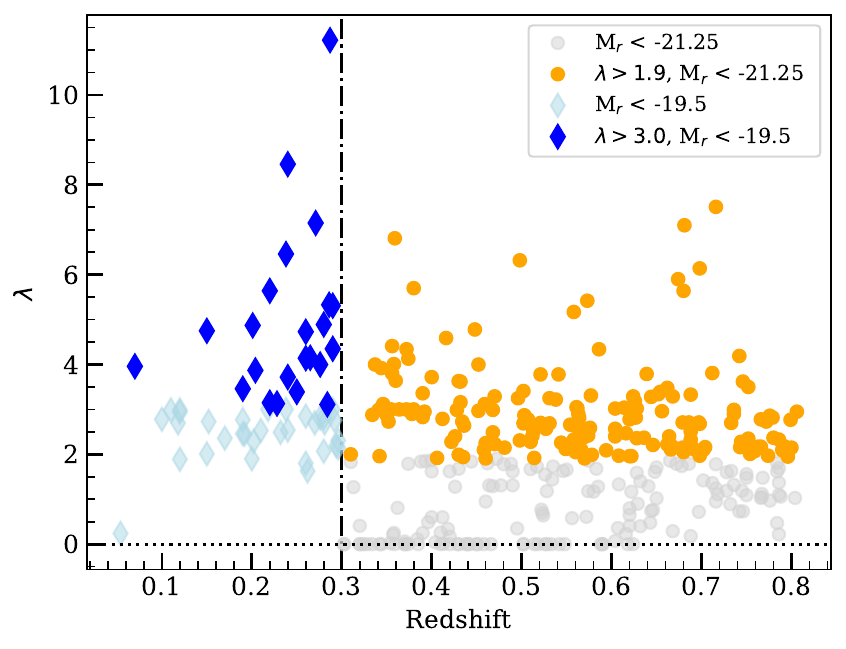}
    \caption{Richness distribution of the cluster candidates detected by PZWav for both group (blue) and cluster regimes (orange) - before (lighter) and after (darker colours) the richness threshold. The dashed line indicates the redshift transition at $z=0.3$.}
    \label{fig:pzwav_richness}
\end{figure}

In Figure\,\ref{fig:pzwav_richness} we present the richness distribution as a function of redshift for both group (blue) and cluster regimes (orange). The lighter colours represent the distribution before applying the richness threshold, while the darker ones symbolise the distribution after the threshold is applied. Within the identified structures, there is a specific cluster, mJPC2470-1771, which exhibits a high richness value of $11.5$ (M$_r<-19.5$) at redshift $z=0.29$, and has been studied by \cite{Rodriguez2022}.  

Finally, following the richness selection, the optical properties of our catalogue range from $1.9 < \lambda< 12$ for the richness, $10.6 < {\rm log_{10}(L_\lambda/L_\odot)} < 11.8$ for optical luminosity and $10.6< {\rm log_{10}(M_\lambda^{\star}/M_\odot)} < 12.2$ for stellar masses.

\section{Results} \label{Results}

In this section, we explore and compare the PZWav galaxy cluster catalogue with others previously presented in the literature. We focus on identifying the optical counterparts within the overlapped region of the AEGIS X-ray survey \citep{Erfanianfar2013} and create a new X-ray catalogue based on PZWav centre estimations. With the available masses, we derive scaling relations with the optical proxies provided by AME.     
Furthermore, we compare different optical catalogues produced for the miniJPAS, by VT and AMICO algorithms, which are discussed in more detail in \S\,\ref{sec:vt} and \S\,\ref{sec:AMICO}. We investigate the variations in centralisation and redshift estimations by highlighting the galaxy clusters that exhibit clusters that are identified by the three cluster finders simultaneously. Additionally, we present statistical analyses of the optical proxies for both matched and non-matched structures across these catalogues.
Finally, we evaluate the significance of each optical catalogue by utilising the galaxy data as a reliable estimator.

\subsection{Comparisons to AEGIS X-ray catalogue}

The miniJPAS survey and the AEGIS survey \citep{Davis2007} have an overlapping area, allowing for a comparison between optical and X-ray catalogues. The X-ray catalogue, created by \cite{Erfanianfar2013} using data from the Chandra and XMM-Newton telescopes, consists of $52$ clusters within the redshift range of $0.06<z<1.54$. These clusters and groups present a mass range of $M_{200} \sim 1.34 \times 10^{13} - 1.33 \times 10^{14}$ M$_\odot$. 
The catalogue is constrained to focus our analysis on the specific area of miniJPAS and the desired redshift range ($0.05<z<0.8$), resulting in a selection of $36$ clusters.

To establish a correspondence between the PZWav optical and X-ray catalogues, we perform a matching process based on the closest distances between centres and redshift offset. We apply a maximum centre difference of $0.5$ Mpc and in redshift $\Delta z = 0.05$. 

This matching procedure resulted in $17$ counterparts. From these, $7$ detections are located at low redshift ($z<0.55$) while $10$ are found within the higher redshift range of $0.55>z>0.8$. In Figure\,\ref{fig:mass_dist}, we present the mass distribution as a function of redshift for the X-ray groups represented as open circles. 

Once we have identified the matching counterparts, we can calculate the differences in centralisation and redshift for the matching sub-sample. These differences hold significant value, particularly in gravitational lensing analyses, as they can introduce biases that impact the final results \citep[][]{Parroni2017, Sommer2022}. However, accurately modelling miscentering offsets without a large sample can be challenging. Therefore, this study focuses on observing potential trends rather than providing a comprehensive analysis of miscentering effects.

In general, the median variations found in centre positions between the catalogues are $125\pm60$ kpc. Regarding redshifts, $90\%$ of the matched detections have offsets within $0.02$ for both low and high redshifts. The median values are similar within the comparisons, $0.001 \pm 0.005$ for the entire redshift range. These findings suggest that the chosen matching criteria are appropriate, and increasing the thresholds would not significantly improve the number of successful matches.

\subsection{Extended X-ray catalogue} \label{extended_xray}

To enhance the comparison, we have reanalysed the same Chandra mosaic of the AEGIS field, implementing a lower detection threshold of $3\sigma$ (compared to $4\sigma$ in Erfanianfar et al.). This new analysis takes advantage of Chandra PSF, which removes contaminating point sources to an order of magnitude lower flux. In addition, we added to the mosaic the XMM-Newton observations in overlap with the miniJPAS field. This not only contributes to improved large-scale background subtraction but also proves to be valuable for identifying newly discovered sources.
Given our focus on the cross-matched sample, we established a systematic procedure for X-ray source identification, ensuring the consistency and replicability of our analyses.

We perform a positional match between the PZWav's optical and X-ray sources, allowing a maximum offset of $0.5$ Mpc. Some X-ray sources have more than one optical counterpart. In cases where an X-ray source has multiple counterparts, we employ a threshold-based approach to assess the significance of the matches. Optical groups that are determined to be chance associations with X-ray sources are removed from further analysis.

If an X-ray source still has multiple identifications after this step, it is excluded from the analysis of scaling relations unless one counterpart exhibits a richness that is significantly larger (by a factor exceeding 1.5). This criterion ensures that contamination from another group to the X-ray flux is limited to less than $30\%$, which is lower than the typical statistical error associated with the newly added sources.

For the catalogue of uniquely identified X-ray sources, we compute the rest-frame X-ray luminosity following the same procedures as outlined in \cite{Erfanianfar2013} on aperture to total flux correction and k-correction and the total mass inferred by the weak lensing calibration of \cite{Leauthaud2010}.

The steps resulted in a $37$ X-ray counterpart catalogue, including $20$ detections within the redshift range of $0.05<z<0.55$ and $17$ detections within the $0.55<z<0.8$ range. From those, $20$ are new detections. We report the properties of X-ray counterparts used in this analysis in Table\,\ref{tab:comp_match_xrays}.

\begin{figure}
    \centering
    \includegraphics[width=\linewidth]{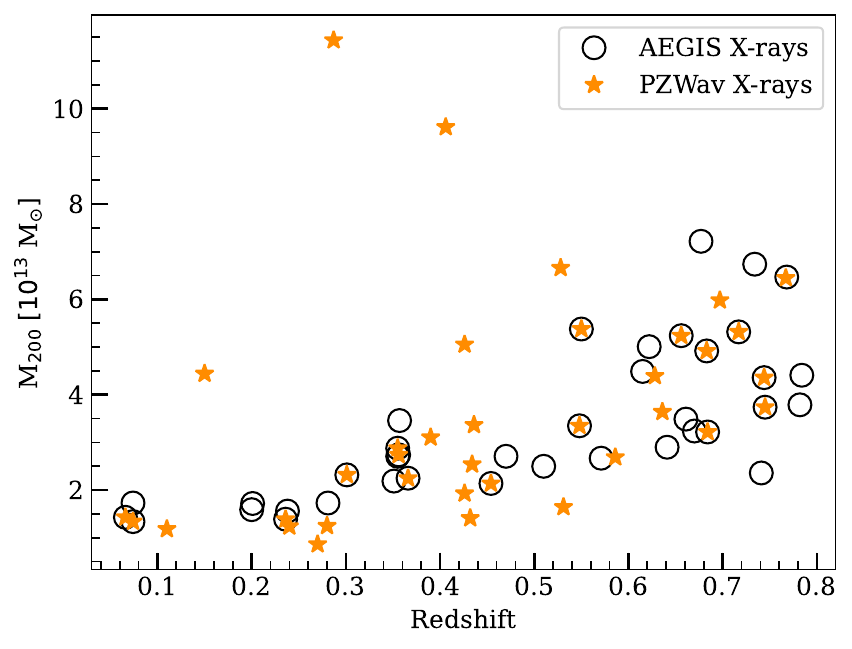}
    \caption{Mass and redshift distributions of the PZWav optical detections that present an X-ray counterpart, as orange stars. Black open circles show Erfaniafar's detection catalogue for the overlapping area. We have $17$ clusters with matching centre distance below $500$ kpc and a redshift offset of $\Delta z < 0.05$. Masses are inferred from \protect\cite{Leauthaud2010} weak lensing calibration.}
    \label{fig:mass_dist}
\end{figure}

In Figure\,\ref{fig:mass_dist}, alongside the X-ray groups identified in Erfaniafar's catalogue (represented by open circles), we show the matched counterparts from the reanalysed sample with the PZWav optical catalogue (indicated by orange stars). The open circles with a central star symbol correspond to the $17$ matches in Erfaniafar's catalogue. The most massive system, located at a redshift of $z=0.29$ and exhibiting the highest richness, has a mass of M$_{200} = (1.14\pm0.07)\times10^{14}$ M$_\odot$. Previous mass estimates based on X-rays alone presented by \cite{Bonoli2021} resulted in $M_{200} = (3.26 \pm 1.4)\times 10^{14}$ M$_\odot$, a slightly larger value but with large uncertainty ($\sim 50\%$). While using Gemini GMOS spectroscopy follow-up, the same authors found $M_{200} = (2.2\pm0.3)\times 10^{14}$ M$_\odot$.

We found a second interesting candidate located at $z=0.41$, close to the survey borders, with an estimated mass of $9.6\times10^{13}$ M$_\odot$. This object was already reported by \cite{Hsieh2005ApJS..158..161H}, as part of the Red-Sequence Cluster Survey \citep{Gladders2005ApJS..157....1G}, with a similar photometric redshift and within $10$ arcsec. As far as we know, there is no dedicated discussion of this object in the literature, and it is a strong candidate for further study.
 
The lowest mass systems range from $(0.72\pm0.31)$ to $(1.60\pm0.24) \times 10^{13}$ M$_\odot$ within the redshift range of $0.07 < z < 0.43$. Additionally, at redshift $z=0.745$, we detect a system with a mass of M$_{200} = (3.75\pm0.5) \times 10^{13}$ M$_\odot$ \citep[also reported by][at similar redshift and within $10$ arcsec.]{Hsieh2005ApJS..158..161H}.

We stress that the successful detection of low-mass groups in our analysis highlights the depth of the miniJPAS survey and demonstrates the effectiveness of our methodology, as also confirmed by the analysis of \cite{Maturi2023}.

\subsection{Scaling relations}

\begin{figure}
    \centering
    \includegraphics[width=\linewidth]{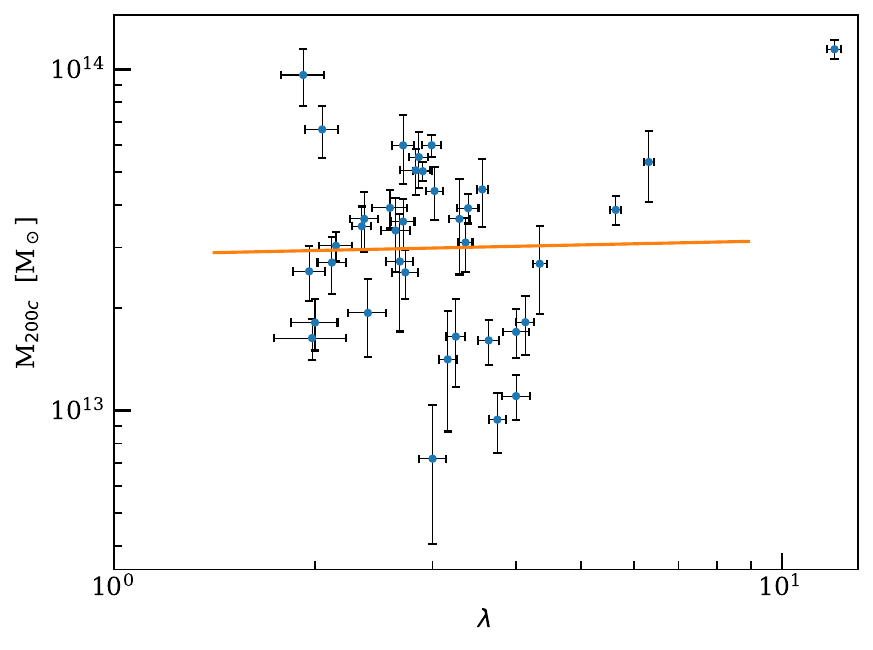}
    \includegraphics[width=\linewidth]{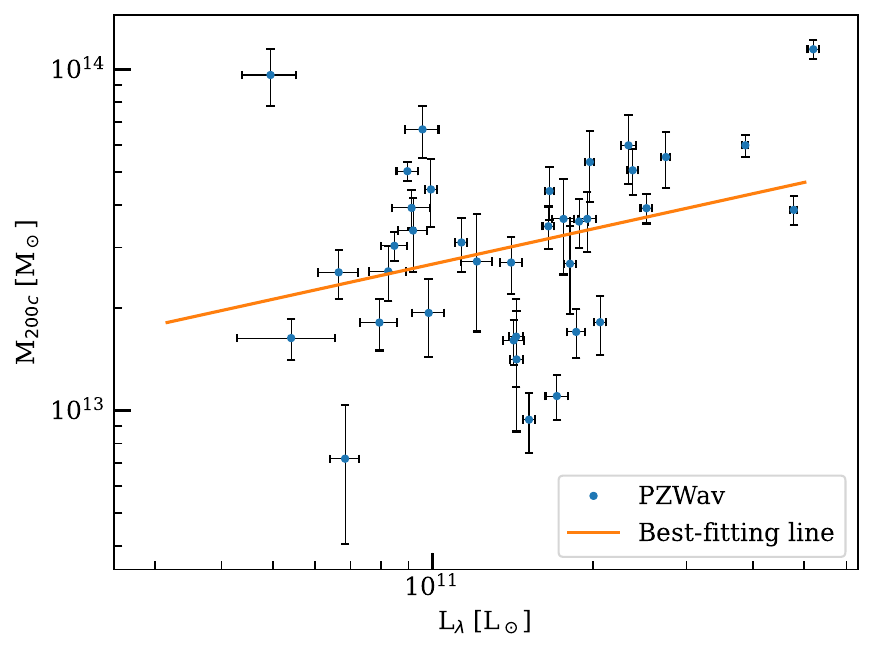}
    \includegraphics[width=\linewidth]{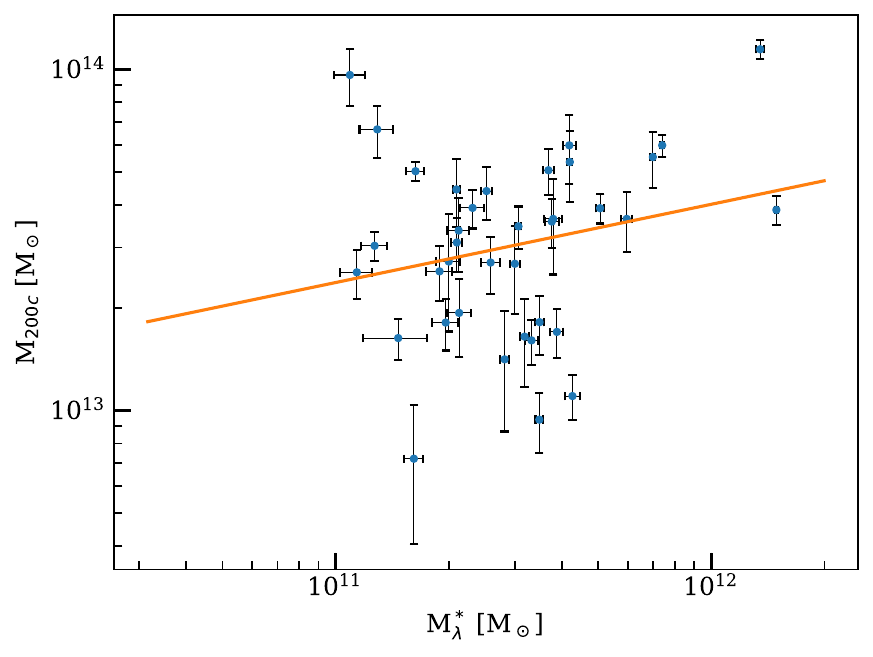}

    \caption{Top panel: M$_{200c}$ estimated from the X-ray luminosity and richness . Middle panel:  M$_{200c}$ vs optical luminosity, $L_\lambda$. Bottom panel: M$_{200c}$ vs stellar mass, $M^{\star}_\lambda$. The orange lines represent the best-fitting values. Parameters can be found in Table\,\ref{tab:fit_values}.}

    \label{fig:M200_qtd}
\end{figure}

Using { \sc LINMIX}, a linear regression procedure with Bayesian approach \citep{Kelly2007}, we are able to perform the minimisation process considering the errors in both the X-ray mass estimates and the mass proxies. The relation is modelled as, 

\begin{equation} \label{eq:reg}
    {\rm log}_{10}\left( \frac{M_{200c}}{{\rm M}_\odot} \right) = \alpha + \beta\,{\rm log}_{10}\left( \frac{O}{\bar{O}} \right) + \epsilon \,,
\end{equation}
where $\alpha$ and $\beta$ are the coefficients, $O$ is the mass proxy, $\bar{O}$ is the pivot value, and $\epsilon$ the intrinsic random scatter about the regression. The best-fitting parameters are given in Table\,\ref{tab:fit_values}, and results are summarised in Figure\,\ref{fig:M200_qtd}. With our sample, we could not find a strong correlation between the quantities. When comparing the mass proxies, it is observed that the richness exhibits the highest scatter for the slope. This behaviour is further accentuated by the limited range of richness values in the sample, as well as the presence of two massive systems with a low richness that stands out from the overall distribution. In contrast, the optical luminosity shows a relatively steep slope, although still within the error bars.

We stress that our sample size is relatively small, which limits the robustness of our statistical results. However, with the ongoing J-PAS survey, we anticipate a larger sample size that will provide more solid and reliable scaling relations.

Improving the scaling relations poses challenges, particularly in terms of mass estimation. The deep X-ray data available for miniJPAS allows us to derive masses at the lower end of the spectrum, reaching a flux limit of $10^{-15}$ erg\,s$^{-1}$\,cm$^{-2}$. In comparison, other surveys such as the ROSAT all-sky survey and eRosita \citep{Merloni2012} have flux limits of 10$^{-13}$/10$^{-12}$ and 10$^{-14}$ erg\,s$^{-1}$\,cm$^{-2}$, respectively.

To address this limitation, \cite{Maturi2023} have proposed alternative approaches. These include reanalysing the XMM-Newton archival data over the J-PAS footprint or incorporating weak gravitational lensing and velocity dispersion measurements into the analysis. These methods offer potential workarounds to improve mass estimation in the presence of flux limitations.

As mass estimates might also be affected by the dynamical state of the structure, filtering the sample to comprise only the most relaxed (~virialized) clusters could improve the scaling relations.

\begin{table}
    \caption{Linear regression fitting values. The mass-observable model is described by Equation\,\ref{eq:reg}. $L_{\lambda}$ and $M^{\star}_{\lambda}$ are given in units of $10^{11}$ L$_\odot$ and $10^{11}$ M$_\odot$.}
    
    \centering
    \begin{tabular}{c c c c c}
    \hline
    Proxy           & $\alpha$         & $\beta$         & $\epsilon$         & $\bar{O}$     \\
    \hline
    $\lambda$       & $13.47 \pm 0.04$ & $0.24 \pm 0.51$ & $0.039 \pm 0.016$ &  $2.9$         \\
    $L_{\lambda}$   & $13.48 \pm 0.05$ & $0.34 \pm 0.18$ & $0.076 \pm 0.019$ & $1.47$\\
    $M_{\lambda}^{\star}$ & $13.48 \pm 0.05$ & $0.23 \pm 0.18$ & $0.080 \pm 0.021$ & $3$\\
    \hline
    \end{tabular}

    \label{tab:fit_values}
\end{table}

\subsection{Comparisons between optical catalogues}

Several computational algorithms can be employed to identify galaxy clusters in optical data, each with its own unique characteristics and capabilities. 
In the following sections, we provide a brief overview of two additional methods, namely VT and AMICO. We present comparative results using the same miniJPAS data set, aiming to understand the differences among the samples detected by all methods or only by a specific method. To assess these variations, we utilise the AME estimator, which allows us to quantitatively analyse the discrepancies between the different catalogues. We emphasise that this study is not intended to highlight the most correct catalogue (pure and/or complete), but rather to provide insights into different properties of the cluster sample, providing an improved analysis.

We stress that there is a dedicated paper from \citep{Maturi2023} on the AMICO catalogue applied in the miniJPAS area.

\subsubsection{VT} \label{sec:vt}

Considering a homogeneous distribution of particles it is possible to define a characteristic volume associated with each particle. This is known as the Voronoi volume, whose radius is of the order of the mean particle separation. Voronoi Tessellation (VT) has been applied to a variety of astronomical problems. A few examples are found in \cite{Ikeuchi1991, Zaninetti1995, El-Ad1996}, and \cite{Doroshkevich1997}. \cite{Ebeling1993} used VT to identify X-ray sources as overdensities in X-ray photon counts. \cite{Ramella2001,Kim2002,Lopes2004}, and \cite{Soares-Santos2011} looked for galaxy clusters using VT. As pointed out by \cite{Ramella2001}, one of the main advantages of employing VT to look for galaxy clusters is that this technique does not distribute the data in bins, nor does it assume a particular source geometry intrinsic to the detection process. The algorithm is thus sensitive to irregular and elongated structures.
 
The parameter of interest in our case is the galaxy density. When applying VT to a galaxy catalogue, each galaxy is considered a seed and has a Voronoi cell associated with it. The area of this cell is interpreted as the effective area a galaxy occupies in the plane. The inverse of this area gives the local density at that point. Galaxy clusters are identified by high-density regions composed of small adjacent cells, i.e., cells small enough to give a density value higher than the chosen density threshold.

To detect galaxy clusters using VT, we use the code employed by \cite{Ramella2001}. It uses the triangle C code by \cite{Shewchuk1996} to generate the tessellation. The algorithm identifies cluster candidates based on two primary criteria. The first is the density threshold, which is used to identify fluctuations as significant overdensities over the background distribution, and it is termed the search confidence level (SCL). The second criterion rejects candidates from the preliminary list using statistics of VT for a Poissonian distribution of particles \citep{Kiang1966}, by computing the probability that an overdensity is a random fluctuation. This is called the rejection confidence level (RCL). More details can be found in \cite{Ramella2001}.

The main drawback of galaxy cluster selection from photometric data is contamination from background and foreground galaxies. A variety of approaches can be applied to deal with this problem. For instance, \cite{Kim2002} used a colour-magnitude relation to divide the galaxy catalogue into separate redshift bins and ran the VT code on each bin. The candidates identified in different bins were cross-correlated to filter out significant overlaps and produce the final catalogue. \cite{Ramella2001} and \cite{Lopes2004} followed a different approach, as they did not have colour information. Instead, they used the object magnitudes to minimise background/foreground contamination and enhance the cluster contrast. The galaxy data was split into overlapping magnitude bins, and the VT code was applied to each magnitude slice. The catalogues of cluster candidates from different layers are combined with a percolation analysis to produce a final list of candidates (see Ramella et al. 2001 and Lopes et al. 2004 for further details).

Taking advantage of the great photometric redshift precision from J-PAS ($0.3\%$), we can divide the input galaxy catalogue into separate photometric redshift slices and run the VT code on each slice. This approach represents a great improvement compared to cluster searches on regular photometric surveys, as it allows a more efficient minimisation of the background.

We ran the VT code on redshift slices of $0.04$ and $0.05$, from $z = 0.01$ to $0.8$. The candidates identified in different slices are cross-correlated to filter out significant overlaps and produce the final catalogue. 

These steps produced a catalogue with $159$ cluster/group candidates.

\subsubsection{AMICO} \label{sec:AMICO}

AMICO \citep[Adaptive  Matched  Identifier  of  Clustered  Objects,][]{Maturi2019}, is a cluster finder based on the search for cluster candidates with a redshift-dependent filter system that seeks to amplify the contrast between the cluster and noise models. It uses the $r$-band for detection as the default magnitude but also accepts one/combinations of other magnitudes. 

The catalogue generated by AMICO with miniJPAS data contains $94$ galaxy clusters for a signal-to-noise ratio larger than $2.5$. It is performed through a redshift range $0.05 < z < 0.8$. More details about the code, such as run parameters, the probabilistic membership association, and the effects of the narrow band photometry on galaxy cluster detection can be found in \cite{Maturi2023}. In this study, the authors derived scaling relations for mass proxies, including amplitude $A$ and estimates of the stellar mass. With a catalogue of cluster members through probabilistic associations, the authors show a good agreement with spectroscopic memberships for galaxies with $P > 0.2$ and identified the Brightest Group Galaxies (BGGs) within the cluster sample. Using AMICO and J-PAS, the study characterised galaxy groups and clusters, down to small groups ($\sim 10^3 M_\odot/h$).

\subsubsection{Matching the catalogues}

\begin{figure*}
    \centering
    
    \includegraphics[width=0.31\linewidth]{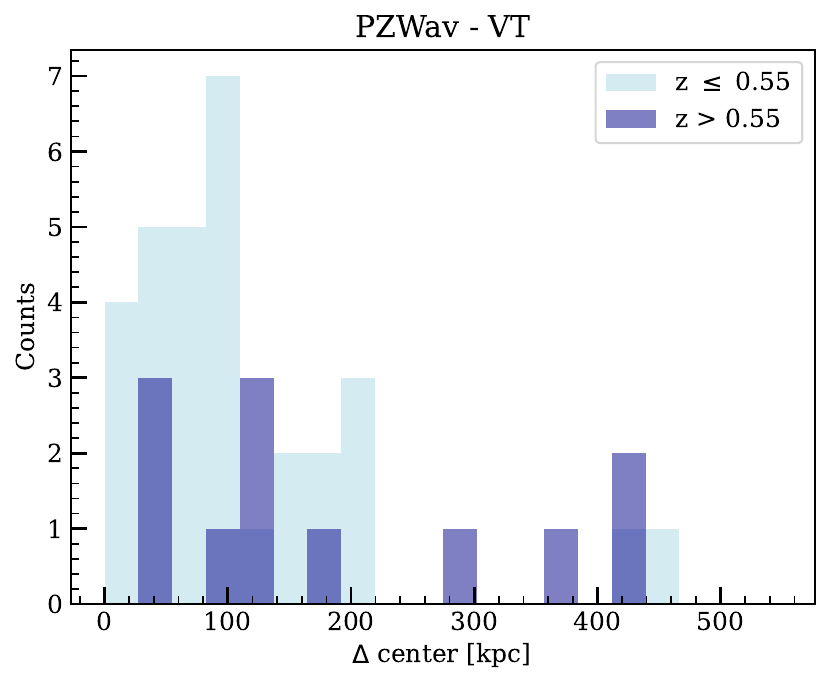}
    \includegraphics[width=0.31\linewidth]{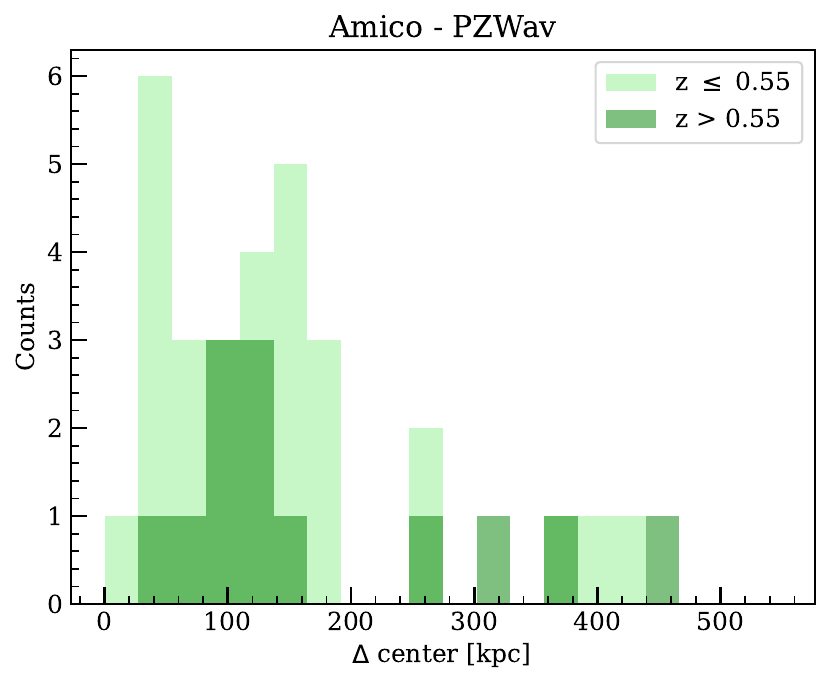}
    \includegraphics[width=0.31\linewidth]{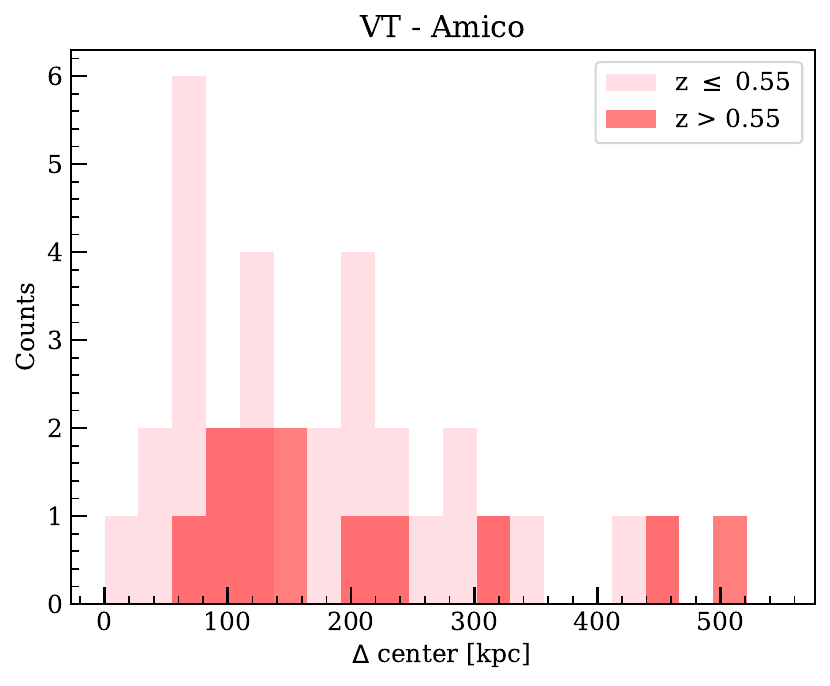}
    
    \caption{Difference between centre coordinates of matching clusters identified by the three optical catalogues, in a comparison 2-by-2. Light colours represent clusters with redshifts lower than $z \leq 0.55$, dark ones for $z > 0.55$. From left to right: Blue: PZWav - VT; Red: AMICO - PZWav; and Green: VT - AMICO.}
    \label{fig:delta_center}
\end{figure*}

\begin{figure*}
    \centering
    \includegraphics[width=0.31\linewidth]{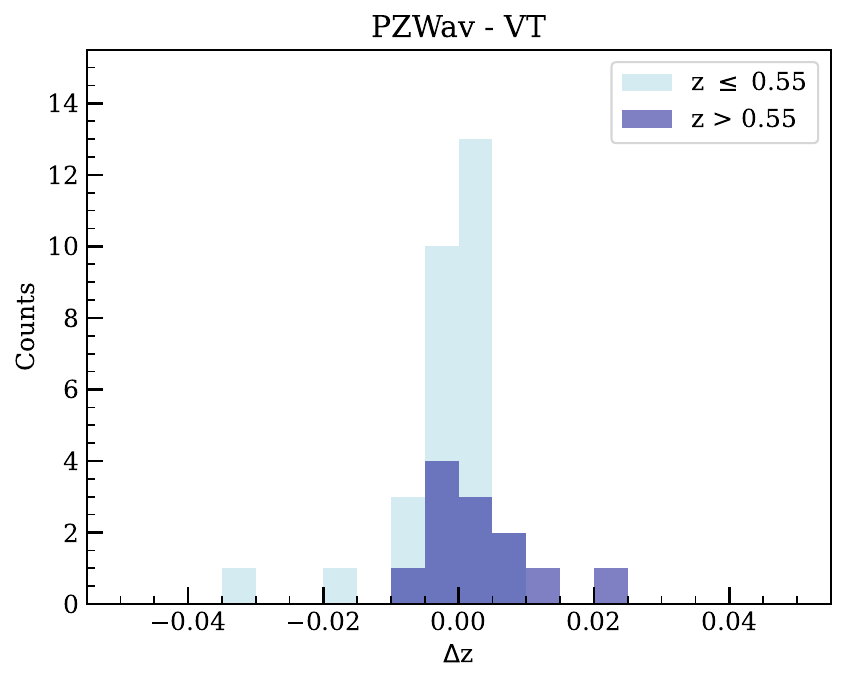}
    \includegraphics[width=0.31\linewidth]{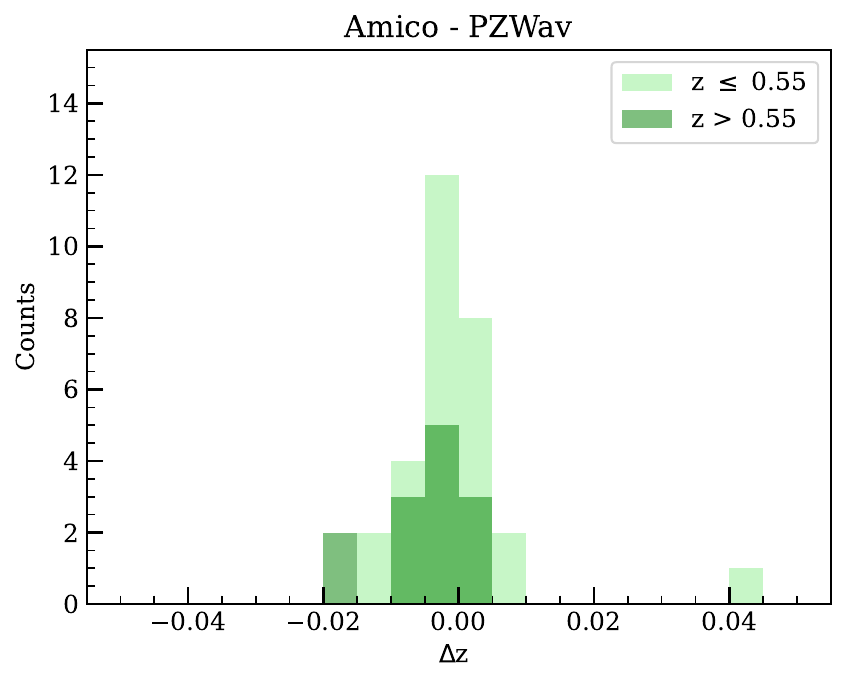}
    \includegraphics[width=0.31\linewidth]{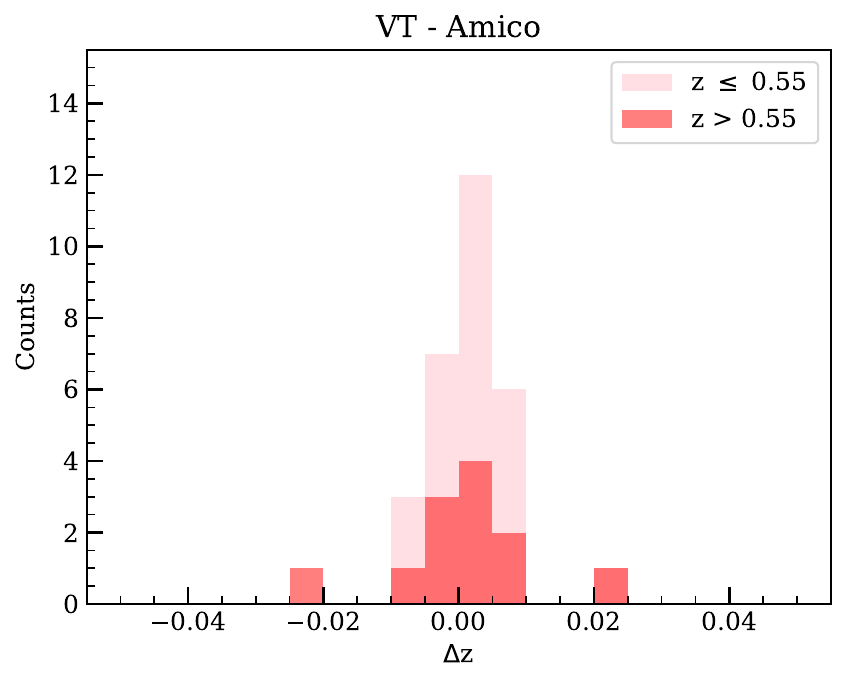}
    \caption{Same as Fig. \ref{fig:delta_center} for the difference between redshifts of the detected structures.}
    \label{fig:delta_z}
\end{figure*}

To compare the PZWav optical catalogue with those produced by AMICO and VT, we keep the same matching parameters of those used in the X-ray comparison, a maximum difference of $0.5$ Mpc between the cluster centres and a redshift difference of $\Delta z = 0.05$. Through this procedure, we identify a total of $43$ clusters that are common to the three cluster finder catalogues. Figures \ref{fig:delta_center} - \ref{fig:delta_z} provide a visual comparison of these matched clusters, showing the agreement between the centre coordinates and redshifts in two-by-two combinations. In these figures, the blue bars represent PZWav - VT comparison, the red bars indicate AMICO - PZWav, and the green bars, VT - AMICO. Figure\,\ref{fig:delta_center} demonstrate that the cluster centres are well characterised, with more than $80\%$ of the clusters with centre differences lower than $300$ kpc. These offsets are lower for low-redshift clusters, with a median value of $92\pm53$ kpc for the PZWav - VT case, compared to the median value of $112\pm67$ kpc for the high-redshift clusters.

Figure\,\ref{fig:delta_z} shows the results in terms of redshift offsets. Here, we observe that more than $90\%$ of the detections show a variation lower than $\pm0.02$, increasing to $95\%$ for high-redshift clusters. The median values of these redshift offsets are similar within the comparisons between the cluster finders, with an overall value of $0.001 \pm 0.005$ for the entire redshift range.
Additionally, both the centre offsets and redshift differences exhibit exponential behaviour, indicating that the chosen matching parameters are appropriate.

\begin{figure}
    \centering
    \includegraphics[width=\linewidth]{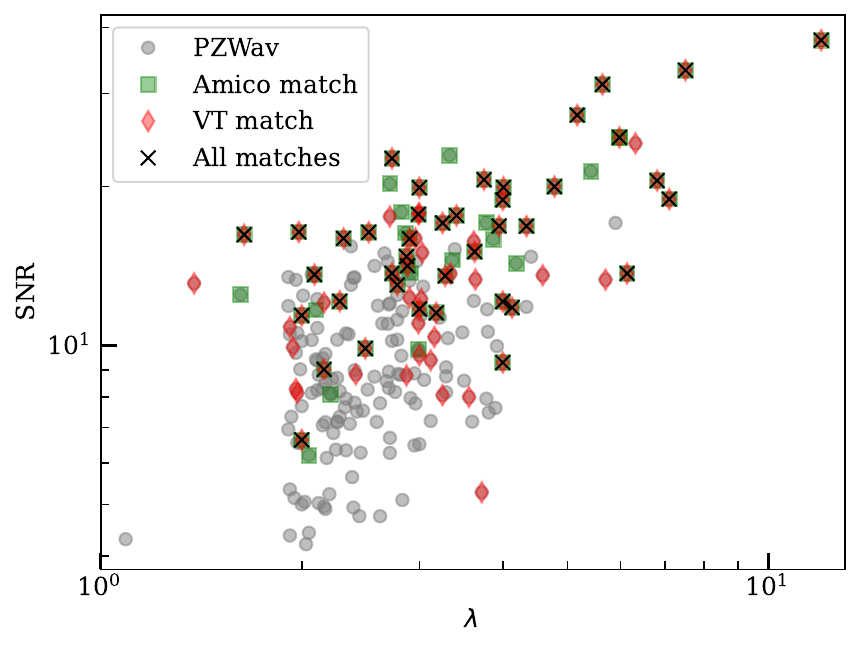}
    \caption{Richness distribution in function of the SNR for the optical catalogues: PZWav in grey dots, PZWav matches to VT in red, and PZWav matches to AMICO catalogue in green. The matches between the three catalogues are presented by ``x''s. The apparent wall at $\lambda=1.9$ is produced by the richness cut described in \S\,\ref{sec:decontamination}. Lower $\lambda$ values are from structures with $z<0.3$, which are affected by another richness threshold.}
    \label{fig:snr_rich_opt}
\end{figure}

In Figure\,\ref{fig:snr_rich_opt}, we examine the distribution of PZWav SNR as a function of the richness (defined as the sum of the galaxies' membership probabilities) in the cluster regime. 
The distribution reveals a correlation between the quantities. When we apply the richness cut, mainly removing structures with low SNR, we observe a decrease in density points for lower SNR values. This behaviour is also evident in the modified shape of the histogram shown in Figure\,\ref{fig:pzwav_detections}. 

We also observe that most matches have a SNR greater than $6$. The common matches among the catalogues are denoted by ``x''s, while the green squares and red diamonds represent the 2-by-2 matches between PZWav and VT, and AMICO, respectively. This analysis shows PZWav's sensitivity to the small density peaks. 

By applying AME on the catalogues, we can provide additional characterisation of the detections using the mass proxies and compare the results with the sample identified by only one of the cluster finders (\S\,\ref{non_match}). Here, we focus on the richness results to calculate the fraction of matched clusters to the total number of detections.

For PZWav, $85\%$ of the catalogue exhibits similarity with the others for clusters with richness ($\lambda$) greater than $6$. This percentage decreases to $30\%$ for richness within the range of 3 < $\lambda$ < 6, and only $13\%$ for $\lambda$ less than 3. Similar trends are observed for VT and AMICO, with $75\%$, $57\%$, $15\%$ for VT and $100\%$, $62\%$, $34\%$ for AMICO, respectively, in the same richness ranges.
These values highlight the agreement between the catalogues.

The significance of the catalogues is further discussed in \S\,\ref{sec:completeness}.

\subsubsection{Statistics on the mass proxies} \label{non_match}

In this subsection, we examine differences between the properties of matched and non-matched galaxy clusters detected by different cluster finders. Table\,\ref{tab:comparison_match_non_match} displays the median values and deviations of richness, optical luminosity, and stellar mass for both the matched and non-matched clusters in the cluster regime.

For the clusters in the matched sample, we observe that the median values show similarities among the different cluster finders, with minor variations within the error bars. This can be attributed to the fact that the matched clusters have similar centre and redshift positions, as they are defined based on small offsets in these parameters. It is important to note that offsets can impact the detection of cluster members. A detailed discussion on how offsets can impact AME's membership detection can be found in \cite{Doubrawa2023}. Here, we highlight the values found for PZWav: $\lambda = 3.18 \pm 0.06$, $L_\lambda =  (1.65 \pm 0.07)\times10^{11}$ L$_\odot$ and $M^{\star}_\lambda = (3.49 \pm 0.14) \times 10^{11}$ M$_\odot$.

However, significant differences arise when analysing the non-matched sample. For example, in the case of PZWav, the richness of the non-matched clusters ($178$) is considerably lower, with a median value of $2.58 \pm 0.03$. This difference is even more pronounced for the VT and AMICO catalogues, where the relative difference exceeds $45\%$. As the PZWav catalogue is limited to $\lambda > 1.9$, the overall offsets are smaller. Similar trends can be observed for the other mass proxies.

We present the median values for the entire redshift range to ensure clarity. However, when examining the sample separately for high ($z > 0.55$) and low ($z < 0.55$) redshifts, the offsets become more significant, with a $75\%$ difference for $L_\lambda$ and an $85\%$ difference for $M_\lambda^{\star}$.

Despite the challenge of distinguishing true detections in observations, this analysis emphasises the median properties of galaxy cluster candidates identified by the cluster finders and are, therefore, likely to be true clusters\footnote{To ensure this conclusion one must rely on simulations.}. Note that the median values of richness for the non-matched detections in VT and AMICO are similar to the richness cut applied during the richness selection of the PZWav catalogue. By redoing the richness analysis, one can suggest that the discussed thresholds in \S\,\ref{sec:decontamination} can be utilised to remove part of the contamination.

It is important to note that the non-matched structures having significantly smaller $\lambda$ values do not indicate a limitation of the detection codes. Instead, it highlights the possible detection of smaller groups and clusters.

\begin{table}
    \caption{Median values of the optical proxies for each cluster catalogue, divided between matched and non-matched clusters, for richness, optical luminosity and stellar mass. M$_\lambda$ is given in units of $10^{11}$ M$_\odot$ and $L_\lambda$ in $10^{11}$ L$_\odot$.}
    \centering
    \begin{tabular}{c c c c c}
    \hline
        & n$^{\rm o}$ cl. & $\lambda$ & $L_\lambda$ & $M_\lambda^{\star}$ \\
    \hline
    PZWav & $43$  & $3.18 \pm 0.06$ & $1.65 \pm 0.07$ & $3.49 \pm 0.14$ \\
          & $178$ & $2.58 \pm 0.03$ & $1.18 \pm 0.02$ & $2.13 \pm 0.07$ \\
    VT  & $43$ & $3.22 \pm 0.07$ & $1.62 \pm 0.07$ & $3.24 \pm 0.18$ \\
        & $116$ & $1.73 \pm 0.12$ & $0.64 \pm 0.06$ & $1.12 \pm 0.10$ \\
    AMICO & $43$  & $3.00 \pm 0.14$ & $1.72 \pm 0.10$ & $3.64 \pm 0.22$ \\
          & $51$ & $1.87 \pm 0.16$ & $0.65 \pm 0.09$ & $1.22 \pm 0.17$ \\
    \hline
    \end{tabular}

    \label{tab:comparison_match_non_match}
\end{table}

\subsection{Significance of the detection catalogues}\label{sec:completeness}

Galaxy clusters are also widely explored in the literature to study cluster galaxy evolution, given the several processes that act on the cluster components through the halo evolution \citep{Evrard1997MNRAS.292..289E, Ettori2009A&A...501...61E, Allen2011, Dvorkin2015}. Thus, understanding the relation between the luminous properties of the galaxies and the dark matter halos ($M_h$), or stellar-to-halo mass relation (SHM), can provide clues about the role of different physical mechanisms that affect the environment. One can quantify the contribution of the brightest cluster/group galaxies (BCGs/BGGs) by constraining a focused SHM relation.
For example, studies by \cite{Leauthaud2012A, Leauthaud2012} and \cite{Gozaliasl2018} provide methods to estimate the halo mass associated with a given BCG based on its stellar mass.

With this relation in mind, we follow the curve derived by \citet{Gozaliasl2018}, and find that galaxies with stellar masses $M^{\star} > 2 \times 10^{11}$ M$_\odot$ are expected to reside in halos with masses of $1.4 \times 10^{13}$ to $1.4 \times 10^{14}$ M$_\odot$. As the SHM relation presents a significant intrinsic scatter \citep[$\sigma_{\rm logM^{\star}} \sim 0.25$, see ][for details]{Gozaliasl2018} the applied limit can not be taken to exclude the presence of less massive galaxies in halos, but as a tool to probe the expected cluster halo masses in our survey.
Although the true distribution of clusters in the survey area is unknown, we can compare the significance of our catalogues with the distribution of galaxies themselves.

To assess the association between galaxies and our cluster/group catalogues, we performed a matching procedure by selecting all galaxies within $1$ Mpc of the optically selected cluster centre and within a redshift range of $\Delta z < 0.05$. Once identified and removed, we can estimate the fraction of galaxies without any association with the catalogues.

\begin{figure}
    \centering
    \includegraphics[width=\linewidth]{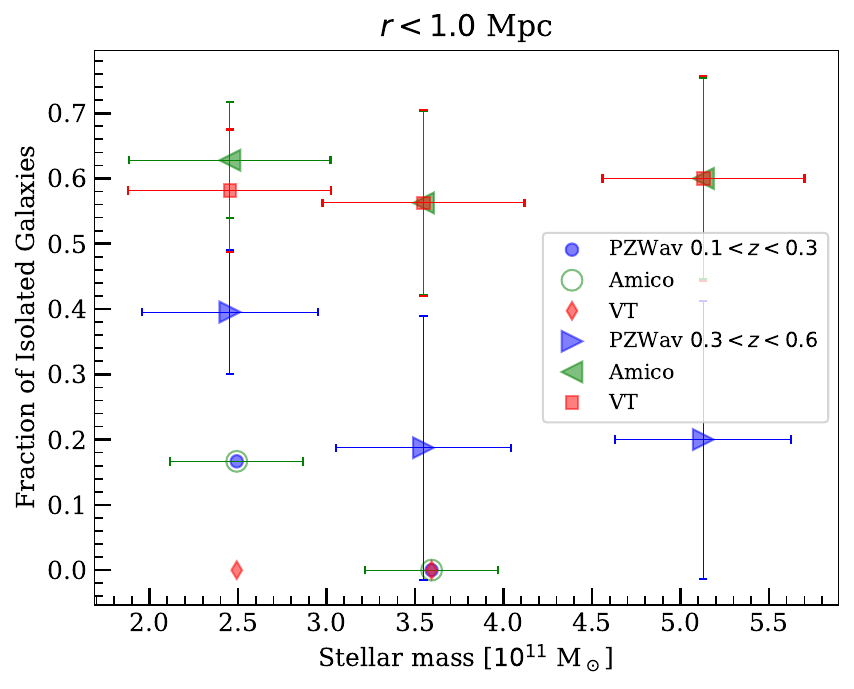}
    \caption{The three optical cluster finders PZWav, AMICO and VT, matched to the generated halo catalogue. The resulting curves give the fraction of galaxies of a given halo mass without any cluster association.}
    \label{fig:match_gal}
\end{figure}

Figure \ref{fig:match_gal} present some conclusions for low ($0.2<z<0.3$) and high redshifts ($0.3<z<0.6$). Note that we do not have results for $z<0.2$, as the catalogue does not present such high-mass galaxies at low redshifts. As PZWav is the largest catalogue, it also has the best matching rates for stellar mass lower than $4\times10^{11}$ M$_\odot$ at high redshifts. In general, we are able to recover approximately $75\%$ of the galaxies. This value increases to $83\%$ for the low-redshift range, indicating that only $17\%$ of the galaxies are isolated. Within $3$ and $5.5\times10^{11}$ M$_\odot$, all galaxies have an associated optical system.  

\begin{figure}
    \centering

    \includegraphics[width=\linewidth]{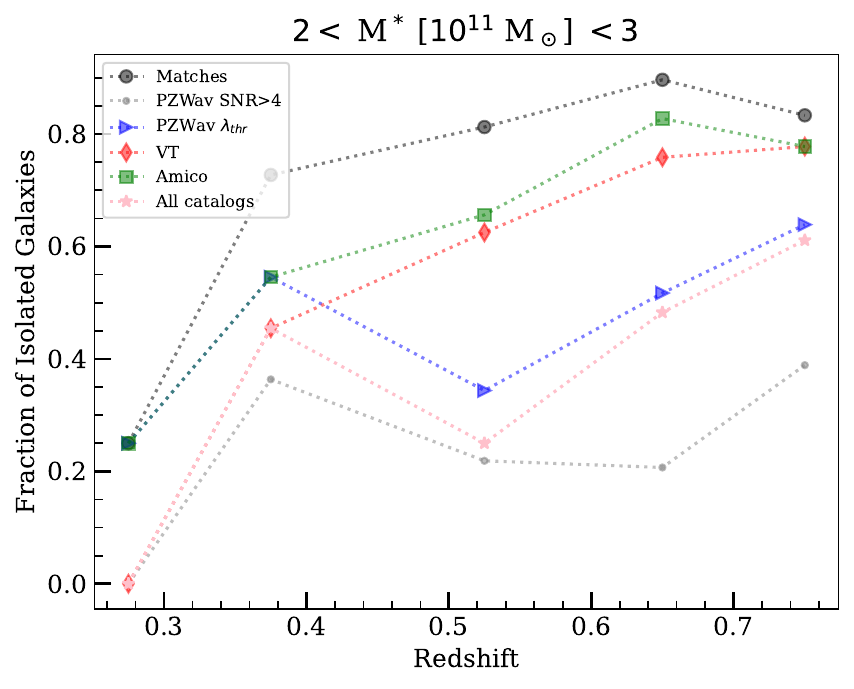}
    \includegraphics[width=\linewidth]{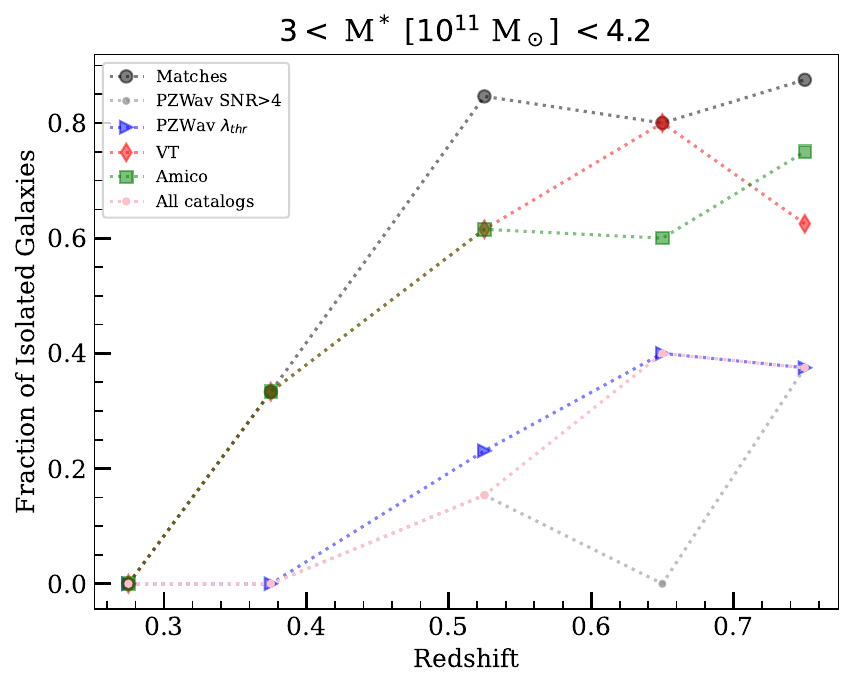}

    \caption{Evolution of the fraction of galaxies without any cluster association with redshift for two mass ranges: $2 <$ $M^{\star}$ [$10^{11}$ M$_\odot$] $< 3$ and $3 <$ $M^{\star}$ [$10^{11}$ M$_\odot$] $< 4.2$, with corresponding halo masses from $1.4$ to $4.1\times 10^{13}$ M$_\odot$ and $0.41$ to $1.4\times 10^{14}$ M$_\odot$.
    In black, the common matches between optical catalogues, grey shows the PZWav complete sample, blue is PZWav after richness selection, red is VT, green is AMICO, and pink is the contribution of all$^{\star}$ optical catalogues together $^{\star}$(PZWav sub-sample).}
    \label{fig:match_gal_mass}
\end{figure}

We analyse the fraction of galaxies associated with optical catalogues as a function of redshift in Figure\,\ref{fig:match_gal_mass} for two stellar mass ranges: $2 <$ $M^{\star}$ [$10^{11}$ M$_\odot$] $< 3$ (corresponding to a halo mass of $1.4$ to $4.1\times 10^{13}$ M$_\odot$) and $3 <$ $M^{\star}$ [$10^{11}$ M$_\odot$] $< 4.2$ ($0.41$ to $1.4\times 10^{14}$ M$_\odot$). The black lines represent the fraction of galaxies that have matches in the three optical catalogues, indicating a more reliable identification, with approximately $20\%$ of galaxies being associated with clusters for $z>0.3$.

In the first mass range, VT (red) and AMICO (green) exhibit similar trends, providing good coverage up to $z=0.3$. PZWav (blue) performs better at higher redshifts ($z>0.5$). The pink curve represents the combined contribution of all the optical catalogues, demonstrating the improvement in the overall fraction rate.
Additionally, the PZWav output with SNR>4 is also shown as a grey curve, as an experiment of the most complete scenario. In this case, only $20\%$ of galaxies do not have optical correspondence.

For the higher stellar mass range, $3 <$ $M^{\star}$ [$10^{11}$ M$_\odot$] $< 4.2$, VT and AMICO exhibit similar behaviour up to $z=0.6$, after which their contributions spread and change positions for $z>0.7$. Due to its larger sample, PZWav shows better fractions across the entire redshift range, with a maximum value of $40\%$ of isolated galaxies for $z>0.65$. 

As an additional analysis, we can assess the significance in terms of the halo mass (M$_h$) by following the relation derived between stellar and halo mass presented by \cite{Gozaliasl2018}. For this, we follow a similar methodology as described above for estimating fractions of isolated galaxies within specific redshift bins while exploring various stellar mass thresholds. Our approach involves adopting redshift bins of width $0.2$ with $0.1$ overlaps and identifying the optimal stellar mass threshold that results in a maximum fraction of isolated galaxies of $0.2$. With the minimum stellar mass results, we derive M$_h$. 

\begin{figure}
    \centering

    \includegraphics[width=\linewidth]{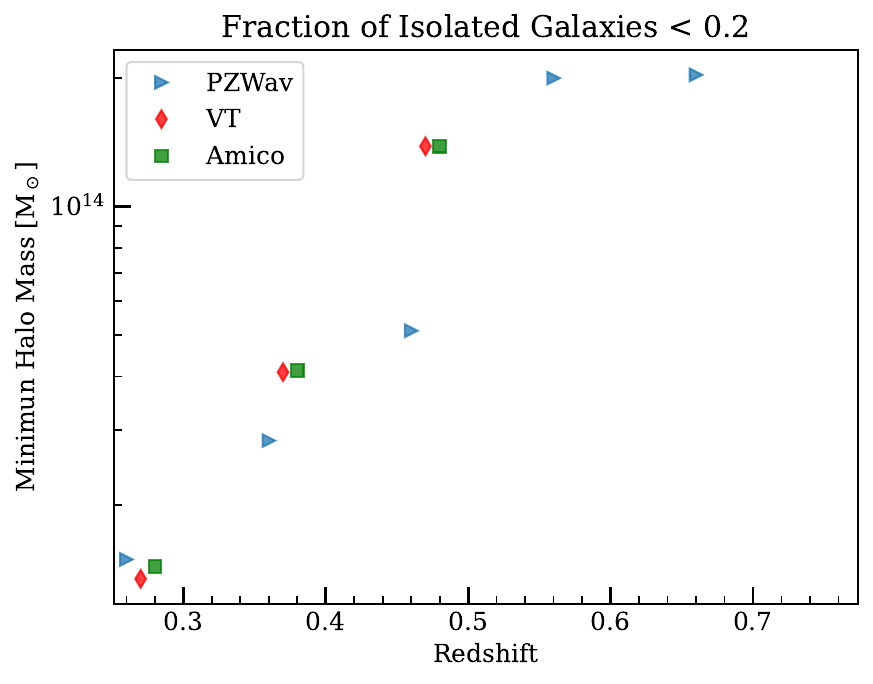}

    \caption{The minimum halo mass in the function of redshift for ensuring a fraction of isolated galaxies lower than $0.2$.
    In blue triangles are presented PZWav $\lambda>\lambda_{\rm thr}$, VT as red diamonds, and green as AMICO values. Halo masses are estimated following the $M^{\star}$-M$_h$ relation described in \protect\cite{Gozaliasl2018}. Mean redshift values are artificially displaced for better visualisation.}
    \label{fig:mhalo}
\end{figure}

In Figure\,\ref{fig:mhalo}, we illustrate M$_h$ as a function of the mean redshift value within each bin for the three cluster finders. PZWav $\lambda>\lambda_{\rm thr}$ is represented by blue triangles, VT by red diamonds, and AMICO by green circles. To enhance visualisation, redshifts are artificially displaced by $0.01$ and $0.02$ for VT and AMICO respectively. The plot shows a similar trend among the cluster finders up to $z = 0.4$, but PZWav exhibits a higher coverage for $z>0.55$ due to the larger number of detections. 

These results highlight the varying capabilities of different optical catalogues in detecting and associating galaxies with clusters.

It is important to note that the recovery fraction and association rates may vary depending on the specific criteria and matching procedures employed in the analysis. However, this analysis provides valuable insights into galaxy clustering properties and spatial distribution within the studied redshift range.

\section{Summary and conclusions} \label{Conclusions}

We conducted a detailed analysis using the PZWav algorithm on the miniJPAS survey, serving as a precursor to the larger ongoing J-PAS survey. To enhance our analysis, we utilise the adaptive membership estimator (AME) that improved the characterisation of the cluster candidates.

In summary, the key findings of this study are as follows:

\begin{itemize}
    \item The PZWav catalogue is composed of $574$ cluster candidates within $0.05<z<0.8$ at SNR>4. 

    \item Defining different absolute magnitude limits allows us to better characterise the systems, in terms of richness, optical luminosity and stellar mass. We apply M$_r<-19.5$ at low redshift ($z<0.3$, group regime) and M$_r<-21.25$ for high redshift ($0.3<z<0.8$, cluster regime).

    \item Applying a richness selection method (limits of $1.9$ and $3.0$ for cluster and group regimes) results in a subset of $221$ candidates.

    \item Making use of AME, we can calculate optical proxies for scaling relations, such as richness, optical luminosity and stellar mass, weighted by the galaxy membership probability. Scaling relations using PZWav results revealed intrinsic scatter of,
    \subitem $\sigma_{{\rm log_{10}}(M|\mathcal{R})} = 0.039 \pm 0.016$
    \subitem $\sigma_{{\rm log_{10}}(M|L_\lambda)} = 0.076 \pm 0.019$ 
    \subitem $\sigma_{{\rm log_{10}}(M|M^{\star}_\lambda)} = 0.080 \pm 0.021$
    
    \item The overlapping with the AEGIS X-ray survey revealed $17$ X-ray sources with an optical counterpart. Among them, we recovered a structure of $M_{200} = (3.75 \pm 0.5) \times 10^{13}$ M$_\odot$ at $z = 0.745$. Reanalysing the data with the optical centres detected by PZWav as the reference, we generated a new X-ray catalogue with $37$ clusters. Within these, $20$ are new detections.

    \item The matching with different optical cluster finders (PZWav, AMICO and VT) revealed $43$ common identifications. Redshift and centre offsets are within $0.001$ and $0.10$ Mpc, respectively. 
    
    \item  Statistics revealed notable differences between the properties of matched and non-matched galaxy clusters. The median values of mass proxies are consistently higher for the matched clusters, indicating their higher Probability of being a true detection. 
    
    \item The analysis of the fraction of isolated galaxies provided an expectation of the catalogue significance. For low mass halos, $f\sim 40\%$, while higher ones present $f\sim 30\%$.
    
\end{itemize}

We conclude that alongside AMICO and VT, PZWav demonstrates strong performance in detecting galaxy clusters through the utilisation of the probability density function (PDF) of photometric redshifts. The ability to detect low-mass structures at high redshifts highlights the depth of the survey. This preliminary analysis, along with other studies focusing on the miniJPAS survey, shows the potential of the J-PAS survey down to the group scales.

\begin{acknowledgements} 

Based on observations made with the JST250 telescope and PathFinder camera for the miniJPAS project at the Observatorio Astrof\'{\i}sico de Javalambre (OAJ), in Teruel, owned, managed, and operated by the Centro de Estudios de F\'{\i}sica del  Cosmos de Arag\'on (CEFCA). We acknowledge the OAJ Data Processing and Archiving Unit (UPAD) for reducing and calibrating the OAJ data used in this work.

Funding for OAJ, UPAD, and CEFCA has been provided by the Governments of Spain and Arag\'on through the Fondo de Inversiones de Teruel; the Arag\'on Government through the Research Groups E96, E103, E16\_17R, and E16\_20R; the Spanish Ministry of Science, Innovation and Universities (MCIU/AEI/FEDER, UE) with grant PGC2018-097585-B-C21; the Spanish Ministry of Economy and Competitiveness (MINECO/FEDER, UE) under AYA2015-66211-C2-1-P, AYA2015-66211-C2-2, AYA2012-30789, and ICTS-2009-14; and European FEDER funding (FCDD10-4E-867, FCDD13-4E-2685).

This work made use of the computing facilities of the Laboratory of Astroinformatics (IAG/USP, NAT/Unicsul), whose purchase was made possible by the Brazilian agency FAPESP (grant 2009/54006-4) and the INCT-A.

L.D. acknowledges the support from the scholarship from the Brazilian federal funding agency Coordenação de Aperfeiçoamento de Pessoal de Nível Superior - Brasil (CAPES). 

E.S.C. acknowledges the support of the funding agencies CNPq (\#309850/2021-5) and FAPESP (\#2023/02709-9). 

P.A.A.L. thanks the support of CNPq (grants 433938/2018-8 and 312460/2021-0) and FAPERJ (grant E-26/200.545/2023).

R.G.D. acknowledge financial support from the State Agency for Research of the Spanish MCIU through the "Center of Excellence Severo Ochoa" award to the Instituto de Astrof\'\i sica de Andaluc\'\i a, CEX2021-001131-S, funded by MCIN/AEI/10.13039/501100011033, and to financial support from the projects PID-2019-109067-GB100 and PID2022-141755NB-I00.

R.A.D. acknowledges support from the Conselho Nacional de Desenvolvimento Científico e Tecnológico - CNPq through BP grant 308105/2018-4, and the Financiadora de Estudos e Projetos - FINEP grants REF. 1217/13 - 01.13.0279.00 and REF 0859/10 - 01.10.0663.00 and also FAPERJ PRONEX grant E-26/110.566/2010 for hardware funding support for the JPAS
project through the National Observatory of Brazil and Centro Brasileiro de Pesquisas Físicas. LM acknowledges support from the grants PRIN-MIUR 2017 WSCC32 and ASI n.2018-23-HH.0.

S.B. acknowledges support from the Spanish Ministerio de Ciencia e Innovación through project PGC2018-097585-B-C22 and the Generalitat Valenciana project PROMETEO/2020/085.

This paper has gone through internal review by the J-PAS collaboration. We would like to thank the internal referees Adi Zitrin and Carlos Hernández-Monteagudo for their helpful comments and suggestions.

\end{acknowledgements}

\bibliographystyle{aa}
\bibliography{optical_detections.bib}

\appendix

\section{X-ray catalogue}

In \S\,\ref{extended_xray} we presented details about the produced extended X-ray catalogue based on the PZWav's optical detection catalogue. In Table\,\ref{tab:comp_match_xrays} we present the $37$ X-rays counterpart catalogue, within the redshift range of $0.55<z<0.8$. We list the cluster identification number from the optical catalogue (ID), the sky coordinates (RA, DEC), the optical redshift estimate (z), obtained flux and estimated error (Flux, eFlux), the X-ray luminosity and error (Lx, eLx), mass estimates and corresponding uncertainties obtained from Lx-Mass scaling relations from \cite{Leauthaud2012} (M200c, eM200c). 

\begin{table*}
    \caption{Extended X-ray catalogue based on the PZWav optical centres, with rows sorted by M$_{200c}$.}
    \centering
    \begin{tabular}{c c c c c c c c c c}
    \hline
     ID &         RA &       DEC &     z &         Flux &        eFlux &           Lx &          eLx &        M200c &       eM200c \\
     16 & 213.80150 & 52.09198 & 0.070 & 1.05e-14 & 8.06e-15 & 2.15e+41 & 1.64e+41 & 7.22e+12 & 3.16e+12 \\
     11 & 214.98826 & 53.11720 & 0.201 & 1.92e-15 & 6.34e-16 & 3.85e+41 & 1.27e+41 & 9.39e+12 & 1.88e+12 \\
     20 & 214.33039 & 52.59097 & 0.238 & 1.77e-15 & 4.35e-16 & 5.20e+41 & 1.27e+41 & 1.10e+13 & 1.66e+12 \\
    153 & 213.89208 & 52.19922 & 0.432 & 8.04e-16 & 5.34e-16 & 1.01e+42 & 6.74e+41 & 1.41e+13 & 5.44e+12 \\
     79 & 215.00479 & 53.10381 & 0.360 & 1.45e-15 & 3.58e-16 & 1.11e+42 & 2.73e+41 & 1.60e+13 & 2.42e+12 \\
     33 & 214.63203 & 52.46455 & 0.280 & 2.47e-15 & 5.56e-16 & 1.01e+42 & 2.28e+41 & 1.62e+13 & 2.25e+12 \\
    270 & 214.26067 & 52.37080 & 0.531 & 6.70e-16 & 3.26e-16 & 1.49e+42 & 7.30e+41 & 1.64e+13 & 4.75e+12 \\
    103 & 215.00902 & 52.97046 & 0.452 & 1.01e-15 & 2.70e-16 & 1.39e+42 & 3.74e+41 & 1.69e+13 & 2.78e+12 \\
    122 & 214.33778 & 52.24765 & 0.290 & 2.74e-15 & 7.69e-16 & 1.21e+42 & 3.41e+41 & 1.80e+13 & 3.10e+12 \\
    115 & 214.18501 & 52.28390 & 0.374 & 1.66e-15 & 5.37e-16 & 1.37e+42 & 4.46e+41 & 1.81e+13 & 3.56e+12 \\
    219 & 213.91066 & 52.04932 & 0.426 & 1.41e-15 & 6.12e-16 & 1.64e+42 & 7.10e+41 & 1.93e+13 & 4.99e+12 \\
      9 & 214.51725 & 52.38455 & 0.434 & 2.14e-15 & 5.68e-16 & 2.54e+42 & 6.76e+41 & 2.53e+13 & 4.12e+12 \\
     81 & 214.44518 & 52.73449 & 0.620 & 1.08e-15 & 3.25e-16 & 3.41e+42 & 1.02e+42 & 2.55e+13 & 4.68e+12 \\
    113 & 213.93670 & 52.12280 & 0.586 & 1.31e-15 & 6.33e-16 & 3.51e+42 & 1.69e+42 & 2.69e+13 & 7.70e+12 \\
    499 & 214.50621 & 52.45121 & 0.758 & 8.23e-16 & 2.55e-16 & 4.63e+42 & 1.43e+42 & 2.71e+13 & 5.12e+12 \\
    241 & 213.69214 & 52.13455 & 0.698 & 9.62e-16 & 6.23e-16 & 4.27e+42 & 2.76e+42 & 2.73e+13 & 1.02e+13 \\
    198 & 214.36356 & 52.54103 & 0.481 & 2.34e-15 & 3.65e-16 & 3.62e+42 & 5.63e+41 & 3.03e+13 & 2.94e+12 \\
     56 & 213.57773 & 52.15029 & 0.390 & 3.64e-15 & 1.06e-15 & 3.27e+42 & 9.55e+41 & 3.10e+13 & 5.53e+12 \\
    133 & 214.76357 & 52.52109 & 0.436 & 3.35e-15 & 1.35e-15 & 3.97e+42 & 1.61e+42 & 3.36e+13 & 8.18e+12 \\
    151 & 214.83700 & 52.89762 & 0.752 & 1.32e-15 & 3.07e-16 & 6.72e+42 & 1.56e+42 & 3.46e+13 & 4.94e+12 \\
    108 & 214.48984 & 52.43453 & 0.686 & 1.63e-15 & 4.31e-16 & 6.39e+42 & 1.68e+42 & 3.57e+13 & 5.78e+12 \\
    230 & 213.95265 & 52.14622 & 0.624 & 1.97e-15 & 1.04e-15 & 5.97e+42 & 3.16e+42 & 3.64e+13 & 1.13e+13 \\
     87 & 215.58267 & 53.36040 & 0.636 & 1.91e-15 & 6.36e-16 & 6.09e+42 & 2.02e+42 & 3.64e+13 & 7.33e+12 \\
      3 & 214.47856 & 52.58452 & 0.680 & 1.89e-15 & 2.98e-16 & 7.16e+42 & 1.12e+42 & 3.87e+13 & 3.79e+12 \\
     46 & 214.76543 & 52.96109 & 0.654 & 2.06e-15 & 3.24e-16 & 7.00e+42 & 1.09e+42 & 3.91e+13 & 3.82e+12 \\
    332 & 214.08193 & 52.24351 & 0.576 & 2.56e-15 & 5.39e-16 & 6.23e+42 & 1.30e+42 & 3.92e+13 & 5.08e+12 \\
     99 & 215.41558 & 53.38501 & 0.628 & 2.68e-15 & 7.74e-16 & 8.06e+42 & 2.32e+42 & 4.39e+13 & 7.74e+12 \\
    273 & 214.94426 & 53.16735 & 0.150 & 4.20e-14 & 1.56e-14 & 4.08e+42 & 1.51e+42 & 4.44e+13 & 9.96e+12 \\
     97 & 214.79143 & 52.61771 & 0.563 & 4.01e-15 & 4.08e-16 & 9.00e+42 & 9.17e+41 & 5.02e+13 & 3.21e+12 \\
    492 & 215.44998 & 52.87806 & 0.420 & 6.84e-15 & 1.68e-15 & 7.32e+42 & 1.80e+42 & 5.05e+13 & 7.64e+12 \\
      7 & 213.31492 & 52.22110 & 0.498 & 5.48e-15 & 2.15e-15 & 8.98e+42 & 3.53e+42 & 5.34e+13 & 1.26e+13 \\
     35 & 215.66653 & 53.36300 & 0.776 & 2.87e-15 & 8.78e-16 & 1.44e+43 & 4.43e+42 & 5.52e+13 & 1.03e+13 \\
     13 & 215.75333 & 53.26884 & 0.697 & 3.84e-15 & 1.45e-15 & 1.45e+43 & 5.49e+42 & 5.98e+13 & 1.36e+13 \\
     23 & 215.14407 & 53.13991 & 0.736 & 3.55e-15 & 4.18e-16 & 1.54e+43 & 1.81e+42 & 5.98e+13 & 4.41e+12 \\
    550 & 215.37229 & 53.46197 & 0.680 & 4.76e-15 & 1.35e-15 & 1.67e+43 & 4.74e+42 & 6.66e+13 & 1.15e+13 \\
    140 & 214.97817 & 53.22724 & 0.406 & 2.04e-14 & 6.35e-15 & 1.95e+43 & 6.09e+42 & 9.61e+13 & 1.82e+13 \\
      1 & 213.61502 & 51.93392 & 0.287 & 5.10e-14 & 5.11e-15 & 2.15e+43 & 2.16e+42 & 1.14e+14 & 7.21e+12 \\
    \hline
    
    \end{tabular}

    \label{tab:comp_match_xrays}
\end{table*}

\end{document}